\documentclass[preprint2]{proto}
\usepackage{times}
\usepackage{amsmath}

\newcommand{\refs}{\par\noindent\hangindent=1pc\hangafter=1}
\newcommand{\eg}{\emph{e.g.},\ }

\begin{document}

\title{\textbf{\LARGE Global Scale Impacts}}

\author{\textbf{Erik Asphaug}}
\affil{\small\em Arizona State University}

\author{\textbf{Gareth Collins}}
\affil{\small\em Imperial College, London}

\author{\textbf{Martin Jutzi}}
\affil{\small\em University of Bern}

\begin{abstract}
\begin{list}{ } {\rightmargin 1in}
%{\leftmargin -1in}
\baselineskip = 11pt
%rule{4.75in}{0.5pt}
%\vskip 1pt
\parindent=1pc
{\small 
Global scale impacts modify the physical or thermal state of a substantial fraction of a target asteroid. Specific effects include accretion, family formation, reshaping, mixing and layering, shock and frictional heating, fragmentation, material compaction, dilatation, stripping of mantle and crust, and seismic degradation. Deciphering the complicated record of global scale impacts, in asteroids and meteorites, will lead us to understand the original planet-forming process and its resultant populations, and their evolution in time as collisions became faster and fewer. We provide a brief overview of these ideas, and an introduction to models.
 \\~\\~\\~}
 
\end{list}
\end{abstract}  

\section{Introduction}

The most important parameter governing the global extent of an impact is the mass ratio of the projectile to the target, $\gamma=M_2/M_1$. In the case of a cratering event this ratio is small, and there is a well-defined geometric locus. Crater scaling then becomes a powerful tool (e.g. Housen et al. 1983) that allows simple analytical approaches to be applied to determine whether an impact `goes global' -- for instance, whether the surface is shaken everywhere to the escape velocity, or whether the target is shattered or melted. 

At the other extreme, as $M_2 \rightarrow M_1$, there is no impact locus, so the mechanics and dynamics are complex (Asphaug 2010). Crater scaling does not apply, even though the impact physics is fundamentally the same. By definition, these similar-sized collisions (SSCs) are global events. Unlike most cratering impacts, they involve substantial downrange or even escaping motion of the projectile $M_2$, depending on the impact velocity $v_{imp}$ and the impact angle $\theta$. 

Impact velocity is the next most important parameter, because of its dynamical and thermodynamical consequences. In order for an asteroid to be eroded in a cratering impact, for example, one projectile mass of material must be ejected to $v_{ej}>v_{esc}$, where $v_{ej}$ is the ejecta velocity and $v_{esc}$ is the escape velocity described below. The ejection velocity in turn scales with $v_{imp}$, so that a cratering projectile has to strike at a few times $v_{esc}$ in a competent rocky target if it is to cause net escape of material, and an order of magnitude faster to cause net mass loss from a highly porous target (Housen and Holsapple 2011). This implies a very different impact evolution depending on porosity. 

Impact velocity also represents a specific collisional kinetic energy $Q=\tfrac{1}{2} (M_1 v_1^2+M_2 v_2^2)/(M_1+M_2)$ where $v_1<v_2$ are the velocities of the target and projectile in the center of mass frame. If the energy is sufficiently intense, $Q>Q^*_S$, then shattering occurs, breaking the solid bonds of the asteroid into pieces no larger than $M_1/2$. If gravitationally bound (ejected at $<v_{esc}$) then shattering produces a rubble pile as defined below; otherwise if fragments are escaping, the result is a collection of new asteroids. This is the classic example of hitting something so hard that you break it. Catastrophic disruption requires a collision of greater energy $Q>Q^*_D>Q^*_S$, to break any solid bonds and also to overcome internal friction (see below), and to exceed the gravitational binding energy, thereby dispersing the fragments to $v_{ej}>v_{esc}$. For massive bodies, intermediate energy collisions $Q^*_D>Q>Q^*_S$ can lead to complicated (altered and reassembled) geologies.

These are idealizations because impact energy is not deposited uniformly inside a target. Much of this chapter is to study how this deposition occurs, and what it does. The impact angle $\theta$ is especially important in this regard, especially for similar colliding masses, since only a limited amount of angular momentum can be accreted in a collision, and because objects of comparable diameter tend to suffer grazing collisions more often than not. And finally, two asteroids of masses $M_1$ and $M_2$ cannot be thought of as colliding in isolation, even if one ignores all the other asteroids and planets: both bodies orbit the Sun and their fragments continue orbiting on intersecting orbits, so their interaction extends long after the original collision.

\subsection{Mass effects}

Generally speaking, impacts slower than $v_{esc}$ cause accretion, and impacts faster than $v_{esc}$ cause erosion, with the specific boundary depending on the impact angle $\theta$. This is effectively the case for cratering impacts and for similar-sized collisions, so we consider the governing parameter $v_{imp}/v_{esc}$. But $v_{imp}$ can be no slower than $v_{esc}$, the impact velocity of two spheres falling from infinity with initial relative velocity $v_{rel}=0$:
\begin{equation}
v_{esc}=\sqrt{2G(M_{1}+M_{2})/(R_{1}+R_{2})}
\end{equation}
where $R_{1}$ and $R_{2}$ are the corresponding radii. As a rule of thumb, $v_{esc}=R_{km}$, in m/s, where $R_{km}$ is the radius of a spherical asteroid in km; this holds exactly true for bulk density $\rho=1.9$ g/cm$^3$. 

For a 2-body encounter in the absence of gas or other drag effects, conservation of energy implies that
\begin{equation}
v_{imp}=\sqrt{v_{rel}^{2}+v_{esc}^{2}}
\end{equation}
Concerning accretion and erosion, whether by cratering or SSC, it is consequently this relative velocity above $v_{esc}$ that matters most:
\begin{equation}
\phi=v_{rel}/v_{esc}
\end{equation}
The magnitude of $\phi$ defines the dynamical `kick' that drives material out of the gravitational potential of the target.  

Dynamical studies of planet formation have tended to ignore the substantial differences between a collision (two objects' radii intersecting) and an accretion. A collision at $v=v_{imp}$ at a separation distance $r<R_1+R_2$ at contact angle $\theta$ is a geophysically and astrophysically complicated event, so that even the slowest merger ($v_{imp}\sim v_{esc}$, $\phi\sim 0$) is an imperfect accretion. Yet most $N$-body simulations to date implicitly assume that whenever two planetary bodies touch each other, they become a single object orbiting the Sun. This is generally an invalid assumption (Agnor et al. 1999). 

Accretion efficiency $\xi$ is defined as the fraction of the projectile mass $M_2$ acquired by the target $M_1$. Perfect accretion $\xi=1$ makes a final merged body $M_F=M_1+M_2$, so in general
\begin{equation}
\xi=(M_F-M_1)/M_2
\end{equation}
Accretion efficiency is always a few percent less than 1, even in $v_{rel}=0$ collisions, because mergers release a substantial fraction of the gravitational binding energy compared to two contacting spheres. Mass is further lost by angular momentum redistribution, flung out by the spiraling, merging protoplanets. Mass loss in otherwise `perfect mergers' can lead to satellite formation (e.g. Canup 2004) and percentage losses of escaping remnants (Asphaug and Reufer 2013). 

In cratering collisions, $\xi(\phi)$ is a relatively smooth function, becoming negative for $\phi\sim 1$, except in the case for highly porous bodies, where $\xi$ may remain positive (accretionary) to velocities tens of $v_{esc}$ in a process of compaction cratering (Housen and Holsapple 2011). SSCs ($\gamma \gtrsim 0.03$) are substantially different because of geometrical effects, and this leads to generally low accretion efficiency. For SSCs accretion efficiency is a sensitive function of velocity $\xi(\phi)$ that starts near $\xi(0)\lesssim1$. At moderately off-axis angles (only $\sim 15^\circ-30^\circ$) SSCs are grazing, in the sense that most of the colliding material does not physically intersect. The colliding bodies can `bounce', with outcomes very sensitive to $\phi$ and $\theta$. 

As reviewed by Asphaug (2010), there is an abrupt transition in accretion efficiency from $\xi(0)\sim 1$, to $\xi\approx 0$ (little mass change to the target, but huge transformation to the impactor) over the velocity range $1.2 \lesssim \phi \lesssim 2-3$. We call these `hit and run' (see chapter by Scott et al. 2015). And finally, erosion and disruption ($\xi< 0$) occur over the range $\phi \gtrsim 2-3$. The exact boundaries of these curves depend on the composition and degree of differentiation of the colliding bodies, as explored in further detail by Stewart and Leinhardt (2012). 

One of the major unknowns of planet formation, is the kind of collision that dominated the accretion of mass by asteroid parent bodies. If it was accretion of myriad much smaller objects (cratering, $\gamma \rightarrow 0$), then effective compaction might facilitate accretion, with highly porous planetesimals acting as `sponges' to sweep up much smaller objects. Truly primitive bodies would contain a record of compaction events (e.g. Belton et al. 2007). If SSCs dominated ($\gamma \rightarrow 1$), then small planetesimals would be unlikely to accrete further once random velocities were excited to faster than $v_{esc}$. Turbulent stirring of the nebula would lead to their disruption (Benz 2000), one of several `size barriers' to primary accretion (Weidenschilling and Cuzzi 2006). 

One possible solution is to bypass these bottlenecks during the gas phase, for example by streaming instabilities followed by pebble accretion as described in the chapter by Johansson et al. (2015). Such models indicate the rapid formation of asteroid progenitors $\sim 100-1000$ km diameter in the presence of the nebula. 
These bodies would be well sorted by mass, so that their further collisional interaction would be dominated by SSCs. Another implication is that many of these bodies (those containing chondritic abundances of active radiogenic $^{26}$Al) would likely be melted by radionuclide decay during the timeframe of their collisional interaction. The surviving original bodies would include a preponderance of hit and run relics according to Asphaug and Reufer (2014), by the attrition of a population that is mostly accreted into growing planets. 

\subsection{Thermodynamical effects}

Impact heating is also sensitively dependent on velocity. In addition to radiogenic heating, shock and frictional heating were common throughout early solar system history, as recorded in meteorites, with shocks being particularly (but not uniquely) effective at  dissipating impact kinetic energy $Q$ into heat (Ahrens and O'Keefe 1994; Melosh 1989). 

Shocks are produced when the impact velocity exceeds the sound speed, $v_{imp}\gtrsim c_s$. Essentially, impact momentum is added to the medium faster than it can be transported away as a pressure wave, leading to a discontinuous `jump' that causes irreversible thermodynamical effects (heating, compression, momentum deposition). Because impacts are generally faster than $v_{esc}$, this means that planet-scale collisions always involve shocks ($v_{esc}$ of tens of km/s), whereas asteroid-scale impacts must be stirred up externally to high random velocity to produce shocks. 

Random velocities in the Main Belt are excited by gravitational interactions with the terrestrial and giant planets, to velocities that are typically $v_{rel}\sim 2-8$ km/s, exceeding $v_{esc}$ by orders of magnitude. Although we have no direct measurements of asteroid sound speed, an upper limit is probably that of the lunar mantle, $c_s \sim 7-8$ km/s. A more representative sound speed may be that measured by Flynn (2005) and others for competent fragments of chondrite meteorites, $c_s \sim 1-3$ km/s. Thus a great deal of shock processing occurs on small bodies in the evolving and modern solar system. But unlike planet-scale collisions that retain their shock-melted materials, the shocked ejecta in asteroid-asteroid collisions are generally lost as interplanetary dust. 

The sound speed is much slower in rubble piles and near-surface regoliths, perhaps only of order $\sim 100$ m/s, so in this crushable zone it is possible to form melted materials that could be retained in place, or at least, be ejected at $<v_{esc}$.  Housen and Holsapple (2011) find that crater ejection velocities in highly porous silicate material can be more than an order of magnitude slower than crater ejection velocities in competent materials. This means that in addition to being more prone to shock melting and compaction melting, the melted material may be less prone to escaping on highly porous bodies.

\subsection{Velocity relationships}

Impact velocity relative to escape velocity, $v_{imp}/v_{esc}$, governs accretion efficiency and its mass effects, while impact velocity relative to sound speed, $v_{imp}/c_s$, governs the formation of shocks and its thermodynamical effects. Accordingly there are four quadrants of global scale impacts, as summarized in Figure 1. 

\begin{figure*}
\epsscale{1.5}
\plotone{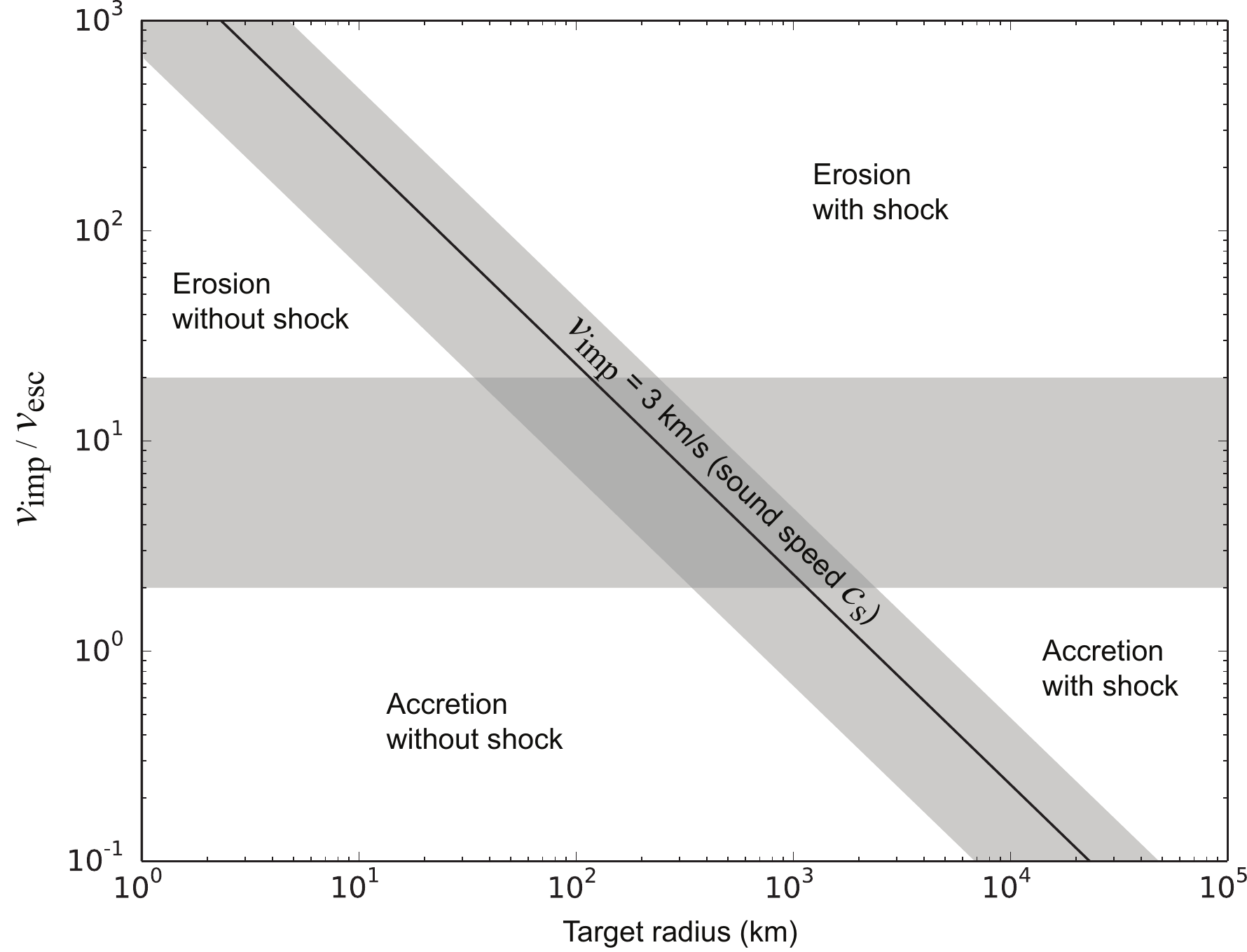}
\caption{\small Global scale impacts span four quadrants in a graph of impact velocity normalized to target escape velocity, $v_{imp}/v_{esc}$, versus target radius. The horizontal band marks the approximate transition from accretionary to erosive/disruptive impacts, a few times $v_{esc}$. The diagonal band defines impact speeds comparable to the sound speeds for a wide range of solar-system materials. Planetesimal-scale accretion (bottom left) does not involve shocks, while planet-scale accretion (bottom right) does. On small bodies, mass loss and disruption do not require shocks. Large $v_{imp}/v_{esc}$ requires gravitational stirring by larger bodies, so that events to the right of the graph tend to be damped (no larger bodies; gravitational drag by smaller bodies) while events to the left tend to be excited, for instance the modern Main Belt.}  
\end{figure*}

Shock-forming impacts are to the right of the diagonal band, represented here by a wide possible range $0.5$ km/s $ < v_{imp} < 8$ km/s. Mass loss and catastrophic disruption ($\xi<0$) occur above the horizontal band, broadly the range $v_{imp} \gtrsim 2$ to $20 v_{esc}$ (competent or highly porous target; cratering or SSC). Asteroid family formation, reviewed in the chapter by Michel et al. (2015), occupies the upper right quadrant, while parent-body accretion occupies the lower left. 

The transitions across quadrants (grey shaded zones) are just as important as the end member scenarios, perhaps moreso. Consider the horizontal band: most collisions will have $v_{imp}$ equal to a few times $v_{esc}$ in a self-stirred population, because the random velocities grow to be proportional to the escape velocities of the largest members (e.g. Safronov 1972). 

As for the sound speed transition, it appears to be coincidence that the range of $c_s$ for asteroid materials overlaps the expected collisional velocities. Random velocities are slower in the outer solar system, $\sim 1-2$ km/s, but so is the sound speed of icy bodies. So in both regions, collisions are expected to occur on either side of the `shock zone', at least in the present solar system. 

During parent body accretion, most events were at or near this line (low velocities), and towards the left (small sizes). As planets grew and stirred things up, events stayed near the line at large sizes, but rose to high above the line for small sizes, grinding down that population into today's generation of asteroids. 

In the first 10-100 Ma after dissipation of the dust and gas in the solar nebula, when collision probabilities were substantially higher than they are today, the average collision speed in the asteroid belt grew from $<1$ km/s to several km/s. Hence, impacts straddled the shock transition zone when impacts were most influential in defining asteroid parent body geology. 

\subsection{Overlapping timescales and linked models}

Global scale impacts involve a diverse range of overlapping processes and timescales, making it important to understand and utilize linked models. Because asteroidal targets vary from tens of meters to hundreds of kilometers diameter, global scale impacts can take place in less than a second, on the ballistic timescale 
\begin{equation}
\tau_{b}\sim 2r/v_{imp}
\end{equation}
where $r$ is the radius of the projectile and $v_{imp}$ is the impact velocity; or in seconds to minutes, on the stress propagation timescale
\begin{equation}
\tau_{s}\sim 2R/c
\end{equation}
where $2R$ is the target diameter and $c$ is the sound speed (or shock speed); or in hours to days, on the gravity timescale 
\begin{equation}
\tau_{g}\sim \sqrt{3\pi/G\rho}
\end{equation}
where $G$ is the gravitational constant and $\rho$ is the asteroid bulk density. 

In the end-member regimes (e.g. self-gravitating planets without strength, or monolithic rocks without gravity) key variables can be eliminated and simple models or scaling predictions can be viable. Similarly, if one can bypass a complex phase or timescale of a collision using an effective approximation, then a simpler analysis can be available -- for instance, skipping the self-gravity phase by applying an escape-energy approximation, or replacing the hypervelocity impact calculation with a `$z$-model' velocity field derived from impact cratering experiments (c.f. Melosh 1989). It is an open question whether these and other simplifications are well justified for modeling global scale asteroid impacts, when the balance of force is so subtle. 

Another approach is to use end-to-end linked numerical models. For asteroids up to tens of km diameter, $\tau_b,\tau_s,\tau_g$ are distinct, making it is possible to consider three separable regimes of impact, using one calculation as input for the next. This of course makes the assumption that not a lot of physics happens in the intermediate regimes, which as we shall see is not always the case.

Consider the example of forming a $\sim10$ km crater on a $\sim20$ km diameter asteroid (e.g. Stickney on the martian satellite Phobos, or Shoemaker on asteroid Eros). Here $\tau_b\sim 10^{-2}$ second, $\tau_s$ is seconds, and $\tau_g$ is hours. An impact hydrocode (Anderson 1987) is a versatile simulation tool (see chapter by Jutzi et al. 2015 and below) that can be run for several $\tau_b$ to compute the shock pressures, energies and temperatures and the initial momentum distribution emplaced by the projectile. The accuracy of this step is limited by our understanding of asteroid target geology, which in even the best codes is represented simplistically (equations of state and crush models). For example, rate dependent effects are not typically included in hydrocodes.

After the impact coupling phase is complete, shock pressures and stress states can be evolved using a wave propagation code (e.g. Richardson et al. 2004; Blitz et al. 2009), or using the hydrocode itself within the limitations and caveats discussed below. Hydrocodes are not generally designed for seismic wave propagation, although as discussed further below, the most modern codes are beginning to do a good job on the equation of state (EOS) aspects of the problem, and the fracture mechanics parts of the problem, and also the granular physics which is intimately connected to the acoustics and flow of disrupted material. Simpler approaches are also feasible, and sometimes recommended, especially those that are anchored in laboratory and field data. A power law of stress attenuation with distance (Cooper and Sauer 1977) can be applied (e.g. Ryan and Melosh 1998; Asphaug 2008), as well as direct comparison with nuclear explosions (Asphaug and Melosh 1993).

If $\tau_g$ is orders of magnitude longer than the other timescales, then the earlier phases of a collision, establishing the pressure evolution, can be calculated with no gravity at all, or with gravity expressed as an overburden term and constant gravity. The largest scale events conclude with a gravitational and rotational-mechanical evolution over several $\tau_g$, as modeled by Michel et al. (2003), Durda et al. (2004), Leinhardt and Stewart (2012) and others; see chapter by Michel et al. (2015).

While gravity can be ignored below some small size, it is not clear what size this is -- gravity being a long-range force and strength, friction, and cohesion being short-range forces. The escape velocity of a 100 m diameter rubble pile is a few cm/s, so massive debris evolution on a small asteroid can perhaps be astrophysically similar to much larger scale events. 

On the other hand, the central pressure inside a 100 m asteroid is only $\sim 1$ Pa (10 dyn cm$^{-2}$), and while not zero, it is hundred times smaller than the cohesions measured for upper lunar regolith. That is, granular cohesion may well dominate over gravity, inside of asteroids up to kilometers diameter, as discussed further in the chapter by Scheeres et al. (2015). 

To make matters worse, granular materials can be cohesive in the absence of internal vibrations, and fluidized by the acoustic energy of a global scale collision, responding like a mobilized global landslide (e.g. Collins et al. 2004). Anyone who has played with sand is familiar with the solid-liquid behavior of granular media. Models are just barely catching up to our awareness of the complex physics that can be at play in a global scale asteroid collision.

\subsection{Small giant impacts}

For large planetary collisions a separation of timescales is generally not possible. In the Moon-forming giant impact, $\tau_b,\tau_s, \tau_g$ are all $\sim 10^3$s, so that all processes must be calculated simultaneously. Fortunately giant impact computations are greatly simplified by two facts: planets can be treated as idealized fluids on these timescales, and the computational timestep $dt$ can be long, requiring fewer hydrocode iterations, as described in Section 2.

For numerical accuracy, $dt$ must be substantially shorter than the wave-crossing-timescale $dx/c_s$ where $dx$ is the spatial resolution and $c_s$ is the sound speed; or in the case of shock-forming impacts, $dx/v_{imp}$. Otherwise information travels faster than the physical propagation velocity in a computation. In medium resolution simulations of the Moon forming giant impact $dt \gtrsim 10$s, which is large enough that an all-in-one 3D hydrocode approach is feasible for the phases leading to the capture of protolunar materials.

Although we don't consider giant impacts in this chapter, we do consider what might be called `small giant impacts', that is, planetesimal-scale SSCs at speeds proportional to the bodies' mutual escape velocity. Like giant impacts, these straddle the erosion-accretion transition in Figure 1. This is thought to be the regime where the parent bodies of asteroids were accreted (see chapter by Johansson et al. 2015). Modeling small giant impacts is most challenging computationally, because of the short timestep combined with complex physics.  Consider a 3D impact simulation of ~10 km planetesimals colliding; $dt < 0.1$s so that it requires $\gtrsim100\times$ more computational effort than a comparable Moon-forming calculation. For well-subsonic collisions the timestep can be sped up by artificially softening the equation of state, thus reducing the sound speed (c.f. Asphaug et al. 2011) while not greatly changing the response so long as the collision remains subsonic.

Although scale-similar to giant impacts, there is usually no shock when the colliding bodies are smaller than about $1000$ km diameter (Asphaug and Reufer 2014). This makes the physics quite different, although there is still substantial heating, shear alteration and melting, all ultimately compensating for the loss of gravitational binding energy. The slowest possible collisions between $\sim100$ km planetesimals occur at $\sim100$ m/s, like two high-speed trains colliding, so friction and compaction do a great amount of work. It is a cutting edge problem in computational modeling, one that can't easily be remedied using linked models on separate timescales, and therefore is in need of experimental work on representative bodies.

\subsection{Rubble piles}

The consequences of accretionary collisions may thus dominate the geologic records of the largest asteroids and primary planetesimals. In small asteroids, on the other hand, this record has probably been beaten down. Because their relative collisional energies are more disruptive, compared to their binding energies (larger $Q$, stirred up by planets), they are second generational or further, as represented by asteroid family formation (chapter by Michel et al. 2015). 

Today most asteroid mutual collisions are erosive or disruptive, $v_{rel}\gg v_{esc}$. Impacts produce craters, sometimes with hemispheric morphologies (mega-craters). If cratering `goes global', as discussed below, the result is global scale resurfacing of the asteroid, by ejecta deposition and by shaking down of pre-existing topography. And of course, impacts contribute great energy of heating, melting, fragmentation, and other forms of material transformation.

Strength is partly a material property, and partly a property of the scale and the dynamics of the collision (Holsapple 2009). Size- and rate-dependent dynamic strength effects (Grady and Kipp 1980; Melosh et al. 1992; Housen and Holsapple 1999) make large bodies much easier to destroy, mechanically, than smaller ones. However, gravitationally large objects are harder to destroy. This means that it is much easier to create a rubble pile, breaking rocks to pieces, than to destroy it, sending half of its mass onto escaping trajectories (reviewed in Davis et al. 2002). In the 1990s it was demonstrated in computer simulations (e.g. Benz and Asphaug 1999) that asteroids larger than a few 100Êm diameter are likely to be pulverized completely, several times over, before they are likely to be destroyed. But as noted by Davison et al. (2013), this does not mean that a given subcatastrophic event is inevitable -- just that comminution is much more likely than disruption, $Q^*_S\ll Q^*_D$. 

According to this analysis, asteroids larger than a few 100 m in a collisionally evolved population become rubble piles. How rapidly or completely they do so, is an open question. It does not mean that asteroids larger than 100 m can be treated as liquid bodies in simulations, or that smaller asteroids are necessarily hard rocks (Holsapple 2009). Indeed it has become obvious in recent years that there is a broad transition regime, rather than a transition size, that spans perhaps $\sim0.1-100$ km diameter, where mechanical, gravitational and impact momentum stresses need to be calculated simultaneously, altogether. This includes all but the very smallest asteroids. 

\subsection{Giant craters}

Giant craters are principal probes of interior geology. They excavate deep inside the asteroid and produce reverberating stresses that cause global surface modification, massive faulting and overturn. Asteroids with giant craters (most of the $\sim10$ km ones evidently) must be `strong' in some sense, in order to survive the mega-cratering events themselves. On Mathilde, several hemisphere-sized simple craters were excavated (or crushed; Housen and Holsapple 1999) without destroying the pre-existing craters, not even their steep walls and rim structures (so far as we can tell at flyby resolution; Figure 2). For an unconsolidated material one might expect sustained reverberations that would trigger an asteroid to relax into a spheroid, but  Mathilde did not; this large asteroid retained some of the most spectacular topography in the solar system. 

To sustain this kind of dramatic topography, one might expect that a solid target structure is required. But this presents a paradox: breaking holes this large into a competent asteroid, will catastrophically disrupt the asteroid, according to strength-scaling rules and hydrocode models with strength (Melosh et al. 1992). So we get backed into a corner, that these large craters cannot have formed in the strength regime. This does not automatically mean they are gravity regime craters, as we shall see.

\begin{figure*}
\epsscale{1.5}
\plotone{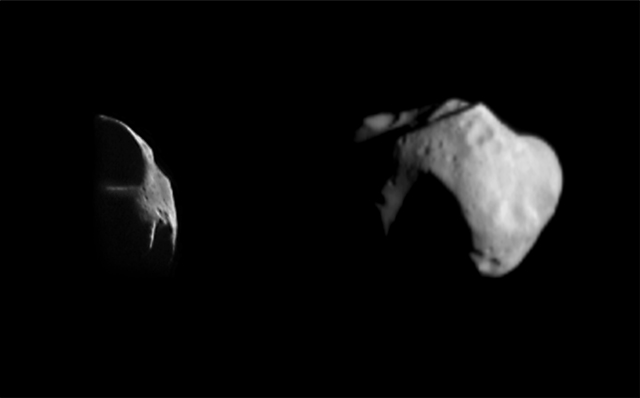}
\caption{\small NASA's NEAR Shoemaker mission obtained flyby images of the ~50 km diameter asteroid 253 Mathilde (Veverka et al. 2000). The image at left was obtained as the spacecraft approached in the direction of the Sun; the image on the right is from greater distance at much lower phase angle ($43^\circ$). Mathilde is a great puzzle for the study of global scale impacts: each giant crater should have reset the asteroid's global shape, according to simulations. Instead each crater seems to have formed in isolation, without degrading the pre-existing craters. If coming in at random angles, these successive global scale impacts would have spun Mathilde into fast rotation, if even a fraction of the incoming momentum was accreted (Asphaug et al. 2002). But the asteroid is one of the slowest rotating bodies in the solar system, $P_{rot}$=418 hrs.}  
\end{figure*}

Of known asteroid craters, only Rheasilivia and Vennenia, the two largest on Vesta (Schenk et al. 2012; Russel et al. 2015, this volume) show clear evidence for gravitational rebound and central peak formation. These two giant basins overlap in Vesta's southern hemisphere, the $\sim$ 500 km diameter Rheasilvia basin overprinting the earlier $\sim$ 400 km Veneneia, offset by some 40$^\circ$, on an asteroid only somewhat larger ($\sim$ 530 km average diameter). Like Mathilde, Vesta provides an example of the interaction of  younger and  older giant craters overlapping, but in this case their interaction is notable. Unlike Mathilde it is easy to recognize the sequence of events, in part because of the much more detailed global imaging but also due to the fundamentally different cratering mechanics on a body ~10 times larger.

A number of past (Asphaug 1997) and recent (Jutzi and Asphaug 2011; Ivanov and Melosh 2013; Bowling et al. 2013; Stickle et al., 2015) numerical impact modeling studies of Vesta have investigated the formation of its mega-craters and their resulting global effects, and the potential for producing ejecta (the V-type asteroids; Binzel and Xu 1993). Modeling two overlapping mega-craters is challenging, and scientifically interesting, because of Vesta's non-spherical shape and gravity potential, rapid rotation, and differentiated structure. In addition, a 3D approach is required, since the problem is strongly non-axisymmetric. 

Jutzi et al. (2013) studied the sequential formation of Veneneia followed by Rheasilvia using a 3D Smooth Particle Hydrodynamics (SPH) impact code described below. The goal of these simulations was to start with a non-rotating differentiated $d = 550$ km sphere, impact it once (to form Veneneia), then place the resulting mega-crater on axis, spin it to period $P_{vesta} \sim$ 5.3 h, and finally impact the rotating cratered body 40$^\circ$ off-axis to end up with the observed topography of Vesta. Thus, Vesta's topography is created by a complex process of ejection and deposition, requiring a realistic treatment of the ejecta gravitational dynamics and the post-impact rheological response. A pressure dependent strength model, a tensile fracture model, a friction model, and self-gravity were included. Finally, to `freeze in' the central peak structures, a  block-model approximation of acoustic fluidization was used. The result (Plate \ref{fig5_3}) is in good agreement with the shape and topography of Vesta (Russell et al. 2015, this volume). 
 
 \begin{figure}[ht!]
 \epsscale{0.82}
 \plotone{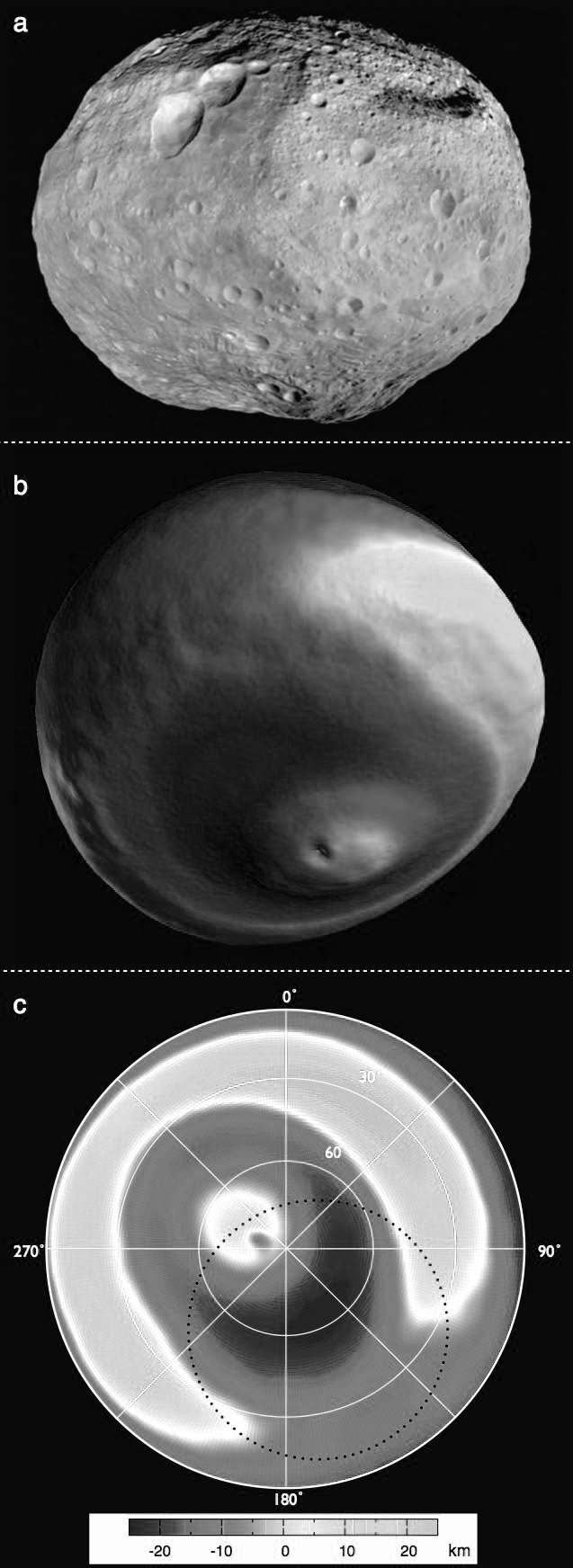}
 \caption{\small a, Asteroid Vesta as seen by Dawn; b, final result of the simulation of two large impacts in the southern hemisphere; and c, Lambert azimuthal projection (equal area) of the southern hemisphere in the model. Shading indicates the elevation (in km) with respect to a reference ellipsoid; for color see Jutzi et al. (2013). Photo credit for a, NASA/JPL/UCLA/MPS/DLR/IDA.}
 \end{figure}
 
Understanding the formation of Vesta's giant craters and the provenance and specific distribution of ejecta is key to understanding the observed properties of Vesta, such as the topography and the surface mineralogy. The proper interpretation of surface materials (and exhumed V-type asteroids and HED-type meteorites) lead to important constraints on models of internal structure (McSween et al. 2013; Clenet et al. 2014). Jutzi et al. (2013) found that a significant fraction of the rocks exposed in the South Pole region should come from $>$50 km deep, and up to $\sim$60-100 km deep in the central mound of Rheasilvia and in the region where the two basins overlap. Also, according to the model, a significant fraction of the material that escapes Vesta during the more recent impact, which would be the dominant source for HEDs, comes from greater than 40 km deep (Clenet et al. 2014).  This is puzzling because olivine-rich mantle rock has not been detected in the Veneneia/Rheasilvia region on Vesta, and there is also a lack of mantle samples among the HED meteorites.

The rapid rotation of Vesta and other asteroids can lead to curious mega-cratering results. In the course of reproducing Rheasilvia on an already-spinning Vesta, Jutzi and Asphaug (2011) found that ejecta from mega-craters would be draped back on the asteroid in complex overlapping lobes, falling back in the rotating frame. Since for Vesta the crater rebound timescale is comparable to the rotation timescale, the Rossby number of the flow $R_o \sim 1$, and the result is actually `Coriolis topography' as shown in Figure 3 -- that is, crater rebound morphologies that form a spiral pattern relative to the impact locus. This provides a potential window to further explore mega-cratering mechanics, especially in materials not covered beneath impact melt  (Schenk et al. 2012).

\begin{figure*}
\epsscale{1.5}
\plotone{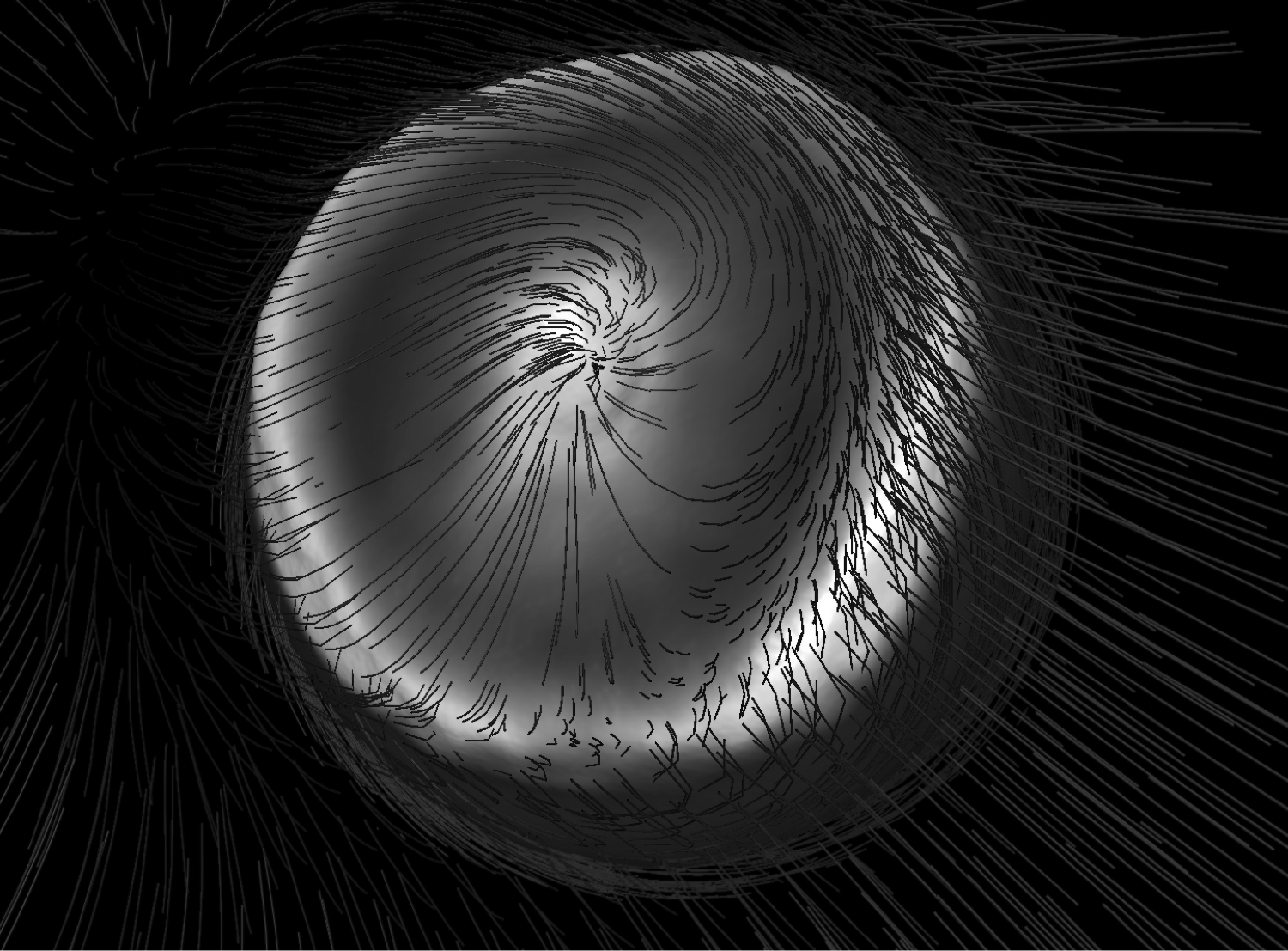}
\caption{\small Simulating the opening and collapse of the Rheasilvia complex crater on asteroid Vesta; from Jutzi et al. (2013). The rapid rotation of the asteroid leads to strong Coriolis forces in the rebounding ejecta, as indicated by the vectors and in agreement with the patterns on the crater floor. Outward vectors plot the distant and escaping ejecta.}  
\end{figure*}

\subsection{Seismic erasure}

There is a strong correlation between the normalized diameter of the largest undegraded crater $D_{max}$ seen on an asteroid, and the diameter $D$ of the asteroid. $D_{max}$ is well defined for most imaged asteroids, and for a number of the best resolved radar-imaged asteroids (see chapter by Benner et al. 2015). Define $\chi=D_{max}/D$. The smallest asteroid seen up close in high detail, Itokawa (mean diameter $D=330$m) has a largest crater $D_{max}\sim 30$ m, that is, $\chi \sim 0.1$. The crater is not obvious to see. For Mathilde ($D\sim 53$km), $\chi \sim 0.6$ -- and is reasonably well determined even from medium-resolution images. For Eros ($D \sim 20$ km), $\chi \sim 0.4$. The trend of $\chi$ varying inversely with $D$ appears to hold generally (Asphaug 2008). Either global spanning craters do not form on small asteroids, or they are unable to retain them.

Shoemaker Crater on Eros shows a different kind of evidence (and so perhaps would Mathilde if imaged at good resolution). The formation of this 7 km recent basin appears to have shaken down regional pre-existing $\sim100$ m scale topography, both next to and opposite the crater (Thomas and Robinson 2005). The seismic erasure may have been impulsive (as seems to be the evidence), taking seconds to minutes as the initial pulse radiated from the impact; or else diffusive as acoustic energy remained trapped inside the asteroid for many $\tau_s$. 

If seismic erasure is efficient, then why has not wholesale seismic degradation shaken down the early giant craters of Mathilde? One approach (Asphaug 2008) is to define a self-consistent critical crater diameter $D_{crit}$ for which the above is true, and to show that $D_{crit}$ increases with asteroid diameter $D$. Craters larger than $D_{crit}$ shake down all pre-existing topography smaller than $D_{crit}$. The giant crater that is forming, continues its excavation long after the seismic impulse attenuates, at least in this scenario, because the crater formation timescale in the gravity regime is many times $\tau_s$. A crater forms as a solitary feature if it has shaken down pre-existing craters of its size. Craters smaller than $D_{crit}$ do not cause global resurfacing at their own scale. Craters larger than $D_{crit}$ stand out as solitary undegraded craters. In this framework one can interpret that $D_{crit}$ has never been achieved on Mathilde (there is no one solitary crater), and that $D_{crit}$ would require something larger than Shoemaker Crater to globally erase the craters of that scale on Eros. 

A normalized critical crater diameter $\chi_{crit}=D_{crit}/D$ can be solved as a function of asteroid diameter and target properties (gravity, density, and scaling constants; c.f. Housen et al. 1983). Assume that the seismic impulse created by an impacting projectile decays in amplitude with distance $r$ from the impact. This impulse is required to shake down topography to a scale equal to the diameter of the crater that is forming, $D_{crit}$, out to a maximum distance $r=D$. If seismic decay with distance takes the form of a simple power law (e.g. Cooper and Sauer 1977), then particle velocity $v_p \propto r^{-\beta}$. Crater diameter $D_{crit}$ is then calculated (Asphaug 2008) according to gravity regime crater scaling, and is substituted into the above, thus solving for $\chi_{crit}$. 

For physically reasonable target parameters, $\chi_{crit}$ is proportional to $D$, accounting for the trend in crater sizes. On a case by case basis, matching the largest undegraded craters observed on spacecraft or radar-imaged asteroids to their theoretical maximum, does a good job at fitting the data if one assumes a seismic attenuation with distance $\beta \sim 1.2-1.3$ for most asteroids. While a power-law attenuation is only an approximation to seismic dissipation, this value is comparable to seismic attenuations of $\sim$0.1-10Êm/s particle velocities in unconsolidated terrestrial materials. 

For Mathilde-sized asteroids, the predicted $D_{crit}$ is larger than the asteroid, consistent with the handful of non-interacting craters greater than half the asteroid diameter. For Itokawa, assuming the same seismic attenuation as Mathilde, it turns out that $\chi_{crit}=0.1$ -- the small crater seen on Itokawa would have resurfaced the asteroid down to that scale. It is an inexpensive if approximate kind of seismology: an Itokawa-sized asteroid exhibing larger craters, would have substantially greater seismic attenuation. The different behavior is in this case not because of material differences, but because of the differences in scale: Itokawa next to Mathilde, is like a refrigerator next to the Empire State Building.

\section{Numerical Modeling of Global Impacts}

When extrapolations lead to curious results, one turns to direct experimentation; when that is unavailable one resorts to the brute force of computer simulations. These techniques are featured in the chapter by Jutzi et al. (2015), and their application to global scale impacts is summarized here. These techniques include hydrocodes (reviewed in Anderson et al. 1987), particle codes (e.g. Richardson et al. 2002), and linked models. These codes rely on models, implemented as part of a time-stepped evolution, describing the material behavior -- such as friction, compaction, viscoelastic properties, and EOS.

Hydrocodes integrate the hydrodynamic equations describing the conservation of mass, momentum, and energy. They relate intrinsic scalar variables -- the density $\rho$, internal energy $U$ and pressure $P$, at position $x,y,z$ -- to the velocity field $v_x, v_y, v_z$ as determined by accelerations (forces). They integrate a time-evolving solution to a set of partial differential equations. If deviatoric stresses are considered, then body forces are evolved according to the stress tensor, introducing six additional variables in 3D; in practice a simplified approach is often made. 

The resulting partial differential equations are integrated forward in time (Benz 1990, Collins et al. 2013) according to accuracy requirements on the spatial resolution and the timestep $dt$. Time-evolution is cast in either a Lagrangian reference frame, that follows the material, or in an Eulerian frame that is fixed in space, advecting material properties (density, momentum, energy, composition) through a grid. These have advantages and disadvantages; for more detailed background and benchmark comparison see Pierazzo et al. (2008). 

\subsection{Smooth particle hydrodynamics}

In smooth particle hydrodynamics (SPH) the continuous fluid is represented as a Lagrangian set of particles (Benz 1990; Monaghan 2012) that move with the flow. The properties at particle $i$ are averaged over its $\sim50$ nearest neighbors, weighted according to their distance from particle $i$ relative to the smoothing length $h$. So $h$ is the grid resolution. Particle density, for example, is the weighted contribution of mass over its neighbors (hence `smoothed').  The mass and momentum conservation equations are approximated to 2nd order as kernel sums. The most commonly used SPH code in asteroid research is by Benz and Asphaug (1994, 1995); this includes elastic strength and fracture damage in addition to the hydro-equations. This code has been modified by Jutzi et al. (2008, 2013) to include friction and compaction. 

The SPH method is mostly applied in 3D, requiring million-particle calculations to resolve colliding bodies $\sim100$ particles across. A fast parallel cluster is needed for production investigations. Nonetheless SPH is popular because it is relatively simple and numerically robust. It is straightforward to add new physics, and the purely Lagrangian description (all variables co-located at particle centers, co-moving with the fluid) makes it easy to hand-off to other codes, visualizations, and analytical reductions. Unlike the grid-based codes described below, the problem domain is completely described by the particles themselves, so there is no need to grid up empty space into which the problem can evolve. 

\subsection{Grid-based codes}

For computation of shocks, and for resolving boundary conditions and abrupt material contrasts, a grid-based calculation generally has greater fidelity than a particle mesh. However, in grid-based Lagrangian hydrocodes mesh-entanglement can result. Eulerian hydrocodes avoid mesh entanglement by maintaining a fixed grid, optimized for the problem. Each time step advects density, internal energy, momentum, and material type from cell to cell, according to the velocity and pressure differential. Material histories and interface positions are recorded using `tracer particles' whose position moves with the mass flux from cell to cell.

Advection of material type is problematic. Imagine iron flowing into a pure olivine cell, whose composition has just become 1\% iron. What EOS do we use in the next timestep? It is not trivial: for instance what if one material is vapor at the computed pressure and temperature, and the other is liquid? There are many approaches to handling mixed materials in a cell. As for the mass flux from a mixed cell, do we assume 1\% iron applies instantaneously everywhere in the cell (faster than any physical velocity), or do we concentrate iron near the donor cell, holding it `upstream' according to some criterion? Eulerian codes usually require techniques for limiting this artificial diffusion.

Grid-based codes have one overwhelming advantage compared to SPH: they allow accurate calculations in 2D (axisymmetric or planar) or 1D (planar or spherical) symmetrical geometries. High definition 2D computations, for instance where a spherical asteroid is impacted on axis, can be done on a modest workstation, enabling careful research exploring many kinds of hypotheses. Parameter exploration in 3D is much more limited. 

One popular grid-based code for asteroid collisional modeling is iSALE (Collins et al., 2004; W\"unnemann et al., 2006), successor to SALE (Amsden et al. 1980, Melosh et al. 1992, Ivanov et al. 1997). This code is maintained in a community forum, and includes both 2D and 3D versions and associated analytical and visualization tools, and several of the most relevant EOS models. Another popular code is CTH (McGlaun et al. 1990, Crawford et al. 2006), although like other sophisticated defense-related packages its foreign access restriction can limit its utility in academic collaborations. 

Which hydrocode to use depends on the problem. An intrinsically axisymmetric problem, such as  impact cratering within about $45^\circ$ of vertical, can be prototyped in a 2D code before serious computer effort is expended looking at the 3D details. If projectile-target contact and compression are of greatest interest, then a 2D or even 1D planar calculation will allow for the highest resolution. If the downrange momentum of massive ejecta is of key interest, then an axisymmetric model at any resolution is almost useless. 

To study the long term self-gravitational interaction following mega-cratering or disruption, the hydrocode simulation can be handed off to a particle method such as we now describe. 

\subsection{Discrete element models}

At low velocity, impacts can be modeled approximately according to Newton's laws using discrete-element methods (DEMs). These do not model shocks or stress waves, but represent collisions by piles of immutable, self-gravitating, idealized solids with contact and restitutive forces (e.g. PKDGRAV; Stadel 2001; Richardson et al. 2002). Soft-spheres codes (Sanchez and Sheeres 2011; see chapter by Murdoch et al. 2015) further represent contact forces according to a viscoelastic 'dashpot' model (not to be confused with SPH). 

Resolution in a DEM does not mean the same thing as it does in a hydrocode. In a hydrocode, the finer the resolution, the more accurate the solution. In a DEM there is a correct resolution for the problem, because every element represents a discrete body. Infinitesimal particle size in a DEM results in infinite strength, and for physical reasons: fine powders are strongly cohesive due to their high specific contact area. In practice, however, DEMs struggle to resolve the physical scale; a $10^6$ element simulation of Itokawa would have a finest block size of several meters. While sufficient to mimic the granular nature of a rubble pile, the contact area per unit volume is a billion times less than if we had a finest block size of centimeters, meaning that we must represent the surface physics parametrically. 

Polyhedral rubble piles codes (Korycansky and Asphaug 2006; Movshovitz et al. 2012) are more challenging to implement than spheres-based DEMs due to the complexity of contact detection and the variety of contacts (face, vertex, edge), and have even worse resolution. But the physical insights are important. Polyhedral DEMs exhibit substantially different behavior than sphere-based DEMs because of the dilatancy and grain locking effects that are only partially captured by spheres-based approaches. For example, Korycansky and Asphaug (2009) find a systematically higher value of $Q^*_D$ in planetesimal collisions at tens of m/s, compared with results derived using spheres-based DEMs (Leinhardt and Stewart 2009). 

For collisions faster than $\sim$10 m/s, DEMs are not appropriate without special handling, because they do not account for realistic mechanical behavior and thermodynamics effects. Comparable to a train wreck, or the collapse of a skyscraper, large planetesimal collisions are transformative events involving fragmentation, crushing, shear friction, and shocks -- they are not piles of bouncing marbles. 

Although the science is still in its early stages, a linked model (e.g. Michel et al. 2003; Leinhardt and Stewart 2012) has proven versatile, where SPH (or another hydrocode) evolves the high velocity collision for a few $\tau_s$, which is handed off as particle positions and velocities to a DEM. But even if the hand-off is perfectly achieved, we caution, as shown below, that the result may depend on unknown aspects of asteroid dynamical rheology. 

\subsection{Strength and damage}

Strength describes a material's response to volume-conserving deformation and to tensile pressure. Volumetric compaction is included in the EOS and the porosity model, described below. The strength model can be simple, requiring no material history, or can be a complicated evolution of strain, strain rate, temperature and pressure.

Asteroid strength modeling used to be based around concepts of brittle fragmentation and fracture damage. While these approaches are still valuable, asteroid strength modeling now revolves more around concepts of cohesion, friction and compaction -- the subtle strengths of granular rubble piles and regoliths, in microgravity.

A small strength or cohesion can exert a major influence on the asteroid impact process, especially during the final stages that govern how the feature will appear geologically, in spacecraft images. Cohesion as small as that of fresh dry snow can dominate global scale events on $\sim100$ m asteroids, making the difference between a cratering event `going global' or remaining local.

A shear strength model dependent on confining pressure $P$ is employed in iSALE (Collins et al. 2004) and in SPH (Jutzi et al. 2013, Jutzi, 2014). In this simple approach the shear stress is limited by a yield stress, which increases with $P$ due to grain interlocking and friction:
\begin{equation}
Y_i = Y_0+ \frac{\mu_i P}{1+\mu_i P/(Y_M-Y_0)},
\label{eq:Yi}
\end{equation}
where $Y_0$ is the shear strength at $P=0$ (measured in the lab), $Y_M$ is the asymptotic shear strength at infinite pressure (the Hugoniot elastic limit), and $\mu_i $ is the coefficient of internal friction. 

In elastic solid materials, a damage $D$ can be accumulated either as a tensor $D_{ij}$ or a scalar (isotropic damage). Damage reduces the deviatoric stress, leaving behind (when $D=1$) only the granular cohesive properties. To account for thermal softening, $Y$ may be further reduced by a factor $1-U/U_{melt}$ where $U_{melt}$ the specific melting energy, so that a molten material has zero strength.  

An approach used in iSALE is to accumulate shear damage as a non-linear function of the equivalent plastic strain (Collins et al., 2004). There is no rate dependence to this function; rate-dependent fragmentation requires a dynamic fracture model (Melosh et al. 1992; Benz and Asphaug 1994, 1995). 

Completely damaged material has some cohesion; lunar regolith has $Y_0\sim 10^3$ Pa. Cohesion of only tens of Pa can dominate the structural geology of small asteroids (Asphaug 2009; Holsapple 2009; Sanchez and Scheeres 2011). This cohesion can be further weakened by granular dynamical processes as discussed in the chapter by Jutzi et al. (2015) and by Kenkmann et al. (2013). 

Acoustic fluidization has been simulated using a block model (Collins 2004) to freeze in the central peaks of the largest craters on Vesta (Jutzi et al. 2013). Although it has not yet been implemented so far for small asteroid collisions, the attenuation of random motions responsible for fluidization might lock in global-scale shapes and structures of rubble pile asteroids.

\subsection{Size- and rate-dependence}

The number of flaws $n$ per unit volume $V$ in a solid is often described as a power law,

\begin{equation}
n(\epsilon)=k\epsilon^{m}
\end{equation}
In this equation a few flaws $n$ will become active in $V$ for a given volumetric strain $\epsilon$. There will be one single weakest flaw, $\epsilon_{min}=kV^{-1/m}$, so that an asteroid's strength decreases with size, as the power $R^{-1/2}$ (Farinella et al. 1982) for $m\sim 6$ (a typical value). This is the static threshold; but dynamic strength increases with strain rate in a closely related manner. The overall effect is a negative slope when plotting $Q^*_S$ versus $R$, where $R$ is asteroid radius -- larger asteroids are weaker. 

On the other hand, larger asteroids are harder to disrupt, because the gravitational binding energy per unit volume (equivalently, the internal hydrostatic pressure $\sim G\rho^{2}R^{2}$) is proportional to the escape velocity squared. Setting gravitational binding energy equal to the rock strength $Y$ gives a transition diameter above which it is more likely for an asteroid to be beaten into rubble many times over, than to be dispersed. 

One of the remarkable discoveries of the previous decade, was that the strength-gravity transition would occur around 100-300 m diameter asteroids. This result was predicted in simple analyses and is confirmed in advanced numerical models, but the implications are still the subject of considerable debate, as reviewed in the chapter by Michel et al. (2015).  

\subsection{Equations of state}

The hydro-equations are closed using an equation of state (EOS) relating energy, density and pressure. For impact modeling, the EOS defines the material response to changes in volume, often in collaboration with a porosity model. The EOS is called by the hydrodynamics integrator (sometimes iteratively) as $P(\rho,U)$ to obtain the `right hand side' terms that drive evolution of the momentum equation.

One popular model is ANEOS (Thompson and Lauson 1972; Melosh 2007), whose thermodynamics variables are derived from the Helmholtz free energy. As a result, all data in an ANEOS table are thermodynamically consistent, something that is especially important in the detailed modeling of shocks. ANEOS returns temperature, pressure, and state of the material, mapping polymorphic and liquid/solid phase transitions, specific heats, and vapor/liquid fraction. But thermodynamical consistency is not the same as accuracy, and Kraus et al. (2012) find that the shock vaporization energy of silicate rocks may be a factor of 2-3 lower than reported by ANEOS, of fundamental importance to the study of high velocity events and giant impacts. 

Another popular EOS model is that of Tillotson (1962), effectively a stitched-together polytrope defining a cold condensed state and a hot expanded state. With only five parameters Tillotson is easily modified to new materials; one avoids the tabular precomputations and interpolations and possible version-inconsistencies that can plague calculations. Tillotson is poorly suited to problems where vapor-melt interactions are expected to dominate, such as the post-impact protolunar `flying magma ocean' (Stevenson 1989). Even so, Canup (2004) finds that Tillotson gives comparable results to more sophisticated EOS models for the initial characteristics of the protolunar disk.

There is a practical advantage to Tillotson: it is simply defined, easily computed, and easily understood in a calculation. Because it derives from the particle-velocity shock-velocity relation (Melosh 1989) it is highly suitable to producing the post-shock particle velocities, and that is what is most important in most asteroid collisions. Melt is not produced in Tillotson (there are no `phases'), but melt and vapor produced in asteroid collisions is usually ejected to escaping velocity anyway, as discussed further below.

\subsection{Porosity} 

The realization that asteroids and planetesimals contain a large volume of pore space (e.g. Housen and Holsapple 1999; Consolmagno and Britt 2002) has driven a recent advance in modeling dissipative properties of porous media (\eg W\"unnemann et al., 2006; Jutzi et al. 2008; Collins et al., 2011). To the extent that resolution allows, macro-porosity can be modeled explicitly using solid materials described above, as cavernous voids (Asphaug et al. 1998; Jutzi et al. 2008). But the idea of a loosely-packed dust rich structure for asteroids (Asphaug 2009) requires porosity to be modeled at the sub-resolution, in addition to the compressibility of a fully-compacted material that is defined by the EOS.

Porosity is associated with a crushing strength. One approach (the P-alpha model; Herrmann, 1969; Carroll and Holt, 1972) computes a distention $\alpha=\rho_s/\rho$, where $\rho$ is the bulk density of the porous material and $\rho_s$ is the density of the non-porous matrix corresponding to the EOS. The pressure $P$ and internal energy $U$ in the porous material, at the grid resolution, is what is needed by the hydrocode, but the EOS only knows how to compute these for the non-porous material, $P_s$ and internal energy $U_s$. 

The P-alpha model uses the approximation
\begin{equation}
     P=\frac{1}{\alpha} P_s(\rho_s,U_s) = \frac{1}{\alpha} P_s(\alpha\rho, U).
\label{meos}
\end{equation}
In this manner the EOS model for non-porous rock is converted into an EOS model for porous rock, whose pressure gradients at the grid scale then drive material accelerations and shocks. The distention varies with pressure according to a `crush curve' derived from uniaxial laboratory experiments, as described in the chapter by Jutzi et al. (2015).

\subsection{Self-gravity}

For modeling global scale impacts into small asteroids, self-gravity can be neglected during the impact phase and applied later, since $\tau_g \gg \tau_s$. Otherwise, self-gravity can be approximated using a constant radial gravity field inside a spherically symmetric asteroid. The modern hydrocodes described above can also calculate gravity explicitly (e.g. Jutzi et al. 2013; Davison et al., 2013), although this adds considerably to the computational expense.

Pre-compression must be included along with self-gravity, otherwise the first phenomenon is the asteroid shrinking and overshooting its central pressure. For small asteroids an analytical pre-compression is acceptable (Asphaug and Melosh 1993). For larger asteroids or more sensitive studies, gravitational stresses must be iteratively solved prior to the impact simulation. 

Following a global scale collision, gravitational evolution must be computed over the orbital timescale of days. This is many thousands of hydro-timescales at the resolution of an asteroid, and thus prohibitive. This evolution is best computed after pressures and velocities have equilibrated, by mapping the hydrocode results into N-body codes or self-gravitating DEMs as described above.

\section{Thermal effects}

Following its accretion in the first few Ma, a typical $\sim$200 km diameter meteorite parent body likely experienced several hundred impacts by asteroids larger than 300 m during the first 100 Ma of solar system history. This is about half the impacts expected during its lifetime (Davison et al., 2013). Over the same period, impact speeds increased from an average $<1$ km/s to as fast as $\sim$10 km/s (O'Brien and Greenberg 2005). 

According to Davison et al. (2013),  post-accretion collisions were only capable of catastrophically destroying at most a few percent of the finished asteroids. Of the remainder, most were subjected to $>1$ collision with an object 1/20 its diameter, and a few percent experienced a collision with an object 1/5 its diameter. 

It is therefore no surprise that evidence for impact processing is common in meteorites (e.g. Sharp and DeCarli, 2006), including shock effects, compaction, and frictional heating. However, as we now review, impact heating appears to be inefficient at the global parent body scale, because the same violent impact that heats and melts the material, ultimately causes the asteroid to disperse ($Q>Q^*_D$). The fate of dispersed shocked material is another question -- whether it finds its home back to the target asteroid, or another asteroid, or becomes ground into fine dust and disappears, or is sampled as meteorites.

\subsection{Global impact heating of asteroids}

Widespread heating and melting of asteroids is evident in meteorites (e.g. Bizzarro et al. 2005) and one of the primary debates has been whether the source of this heating is by early accretion of radiogenic $^{26}$Al, or by impact processes. 

Keil et al. (1997) combined simple theoretical arguments with observational evidence and numerical impact simulations to conclude that the globally averaged effect of impact heating on a non-disrupted asteroid was equivalent to a global temperature increase of less than a few degrees. Others (e.g., Rubin 1995) maintained that shocks from high velocity collisions, especially into porous primitive targets, could be the source of melting, and that the meteorite collection itself would be biased to include the more highly shocked and melted surface materials.     

Davison et al. (2013) revisited the question of impact heating, using a Monte-Carlo simulation that accounted for the effect of asteroid porosity and impact velocity on the disruption threshold (Jutzi et al., 2010; Housen and Holsapple, 1990) and for the evolving impactor size- and velocity-frequency distribution (using a simulation by O'Brien and Greenberg 2005). For a 200-km diameter, 20\% porous asteroid, Davison et al. (2013) found that the maximum specific impact energy that could be imparted, without disrupting the target, is $\sim $0.1-0.7 MJ/kg. This corresponds to a globally-averaged temperature increase of 100-700 K, depending on the impact velocity, substantially greater than the Keil et al. (1997) estimate.

However, an impact just below the catastrophic threshold is not representative of the largest collision that is expected. This just gives the maximum heating that is theoretically possible, without also dispersing the heated mass. Davison et al. (2013) followed the collisional evolution during the first 100 Ma of solar system formation, assuming an approximately inverse square size-frequency distribution of impactors. They found that in most cases, the most energetic impact experienced by a 200-km diameter asteroid that was not disrupted, was only a few percent of $Q^*_D$. They concluded that significant impact heating on a $\sim$200-km diameter parent body is unlikely, and in the end obtained a result similar to Keil et al. (1997), that in the most typical scenario the maximum specific impact energy delivered to a non-disrupted asteroid amounts to a globally-averaged temperature increase of only a few degrees.

Larger asteroids have much greater specific binding energy and can therefore withstand collisions that deposit far more heat. An analogous situation is true of planetary cratering, where the largest craters are the most melt-rich. But a maximal event is less likely to occur on a larger target than on a smaller target, for an approximately inverse square size distribution. On this basis Davison et al. (2013) found that both the maximum specific impact energy and the total impact energy were rather constant for asteroids 100-500 km in diameter, a few tens of degrees of heating at most. These estimates assume that all of the heated materials of impact are retained in the non disrupted asteroid.  A fraction of the hottest material actually escapes the asteroid, so that these global heating estimates are upper limits. 

\subsection{Localized and buried heat}

These globally-averaged estimates suggest that short-lived radionuclide decay ($^{26}$Al) was a far more effective global heat source than impact. However, impact heating is transient, local, and non-uniform. In events where the projectile is much smaller than the target, impact energy is converted into heat within the proximal, near-surface volume that is highly shocked, a few projectile radii, and zero appreciable heating occurs far from the impact.  For small asteroids with low escape velocity, much of this melted material escapes, while on large asteroids it can be retained, requiring a different sort of calculation.

\begin{figure*}
\epsscale{1.5}
\plotone{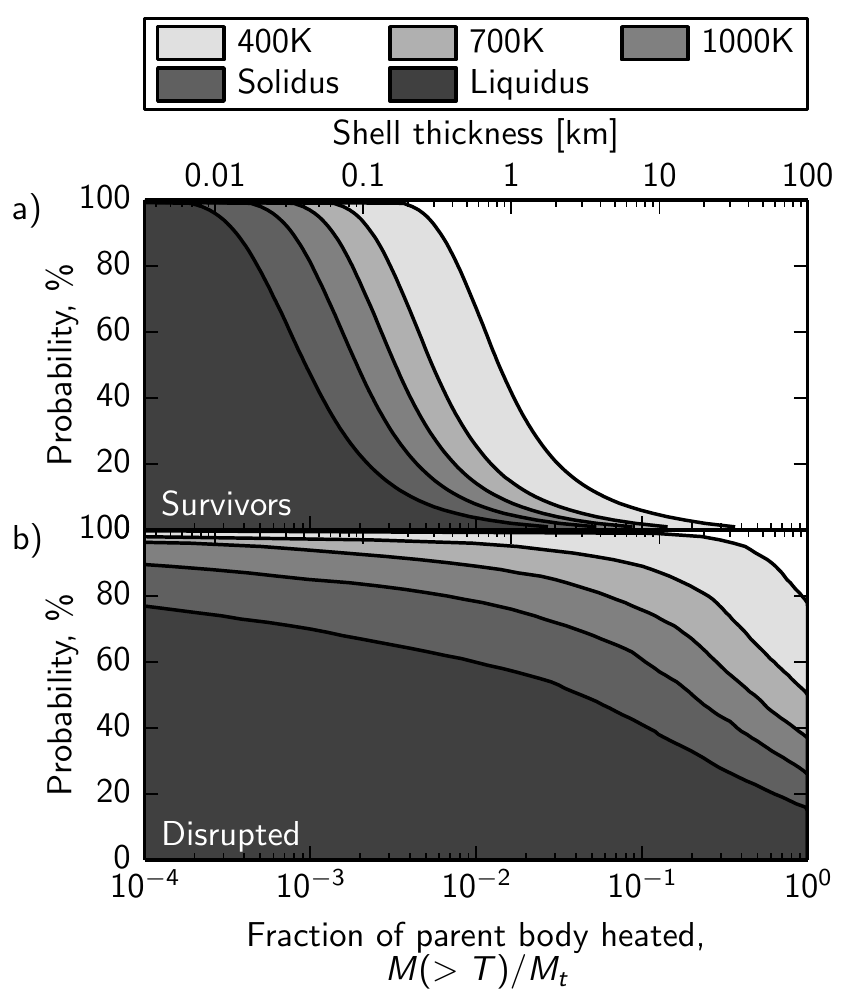}
\caption{\small The probability that a given mass fraction of a meteorite parent body will be heated to a 
range of temperatures, for (a) those asteroids that survive the first 100 Myrs of Solar System history without 
being disrupted and (b) those bodies that are disrupted. Modified from {\em Davison et al.} (2013)}  
\end{figure*}

Temperature increases attenuate with shock pressure, rapidly with distance, and so do ejection velocities. Only in rare, disruptive collisions between objects of similar mass can impact heating be considered to be nearly uniform (Davison et al., 2010). The accretionary collisions described above (`small giant impacts') are global in consequence, but too slow to result in shock heating. Frictional and compaction heating during early accretion (the first $\sim3$ Ma) may be of global consequence, but are thought to be insignificant in magnitude compared to the prevalence of $^{26}$Al.

Using the same dynamical and population framework described above, Davison et al. (2013) computed the volumetric significance of impact heating on 200-km diameter meteorite parent bodies during the first 100 Ma of Solar System history (Figure 4). Less than 1\% by mass of a typical undisrupted parent body experiences heating of more than a few hundred K, volumetrically equivalent to a surface layer $<300$-m thick. However, of the few percent of asteroids that are disrupted, most experience a near-global temperature increase of a few hundred K, and one fifth are expected to undergo (near) complete melting. 

The main influence of asteroid porosity on impact heating is to modify the distribution of heat. Shock heating is more extreme, but much more localized, in porous body collisions compared to impact on a nonporous body. More shock energy is absorbed close to the impact site on a porous asteroid by permanent crushing of pore space; the excess heating from pore collapse leads to higher proximal temperatures (Asphaug et al. 2002; Davison et al., 2010). 

The localized heating by sub-catastrophic collisions, and the distribution of heated materials beneath and around the crater, imply that a single impact can produce materials with a remarkable range in peak temperature and cooling rates. According to Davison et al. (2012), impacts on porous asteroids can produce large amounts of melting at speeds as low as 4 km/s, and material heated up to $\sim$1200 K (petrologic type 6) is expected for most impacts in the first 100 Myrs (Davison et al., 2013). 

Numerical simulations of crater formation in large sub-catastrophic impacts also suggest that heated impactor and target material, and exhumed hot material from the deep mantle, is rapidly buried beneath an insulating lens of porous breccia. The structure of these lobate/hemispheric deposits, for the specific case of the global scale impacts Rheasilvia and Vennenia on asteroid Vesta, is studied by Jutzi and Asphaug (2011) and Jutzi et al. 2013. Buried impact-heated or impact-exhumed material cools slowly, consistent with the wide range of metallographic cooling rates inferred for chondritic materials (Davison et al., 2012). 

While the volumetric significance of impact-heated material appears to be consistent with the limited constraints on meteorite thermal metamorphism, compared to $^{26}$Al decay it appears that localized impact heating is an important secondary effect, even when considering the cumulative effect of multiple impacts over the lifetime of a parent body. There is of course a bias, that the proximity of impact-heated material to the parent body's surface makes it most accessible for excavation and meteorite delivery by subsequent collisions.

\subsection{Accelerated cooling by impact overturning}

If internal heating by short-lived radioactivity was dominant, then sub-catastrophic collisions would have played an additional important role in regulating thermal evolution. Ciesla et al. (2013) showed how a single large impact can have an enormous influence on the cooling of planetesimals heated internally by radioactivity. 

In their models (Figure 5), hot, thermally-softened material is brought up from the deep interior to flow out and over the surface of the planetesimal, turning the asteroid ``inside out." The direct exposure to space of deeply exhumed molten silicates would lead to rapidly accelerated cooling and degassing; more generally, it is a considerable source of enthalpy for materials brought to the surface, something Asphaug et al. (2011) proposed as an origin of chondrules. 

In general, a much larger fraction of an internally-heated planetesimal experienced enhanced cooling than is heated by the impact, implying the nonintuitive result, that the major effect of sub-catastrophic collisions was to accelerate the loss of heat, rather than to deposit heat (Ciesla et al., 2013). 

\begin{figure*}
\epsscale{1.5}
\plotone{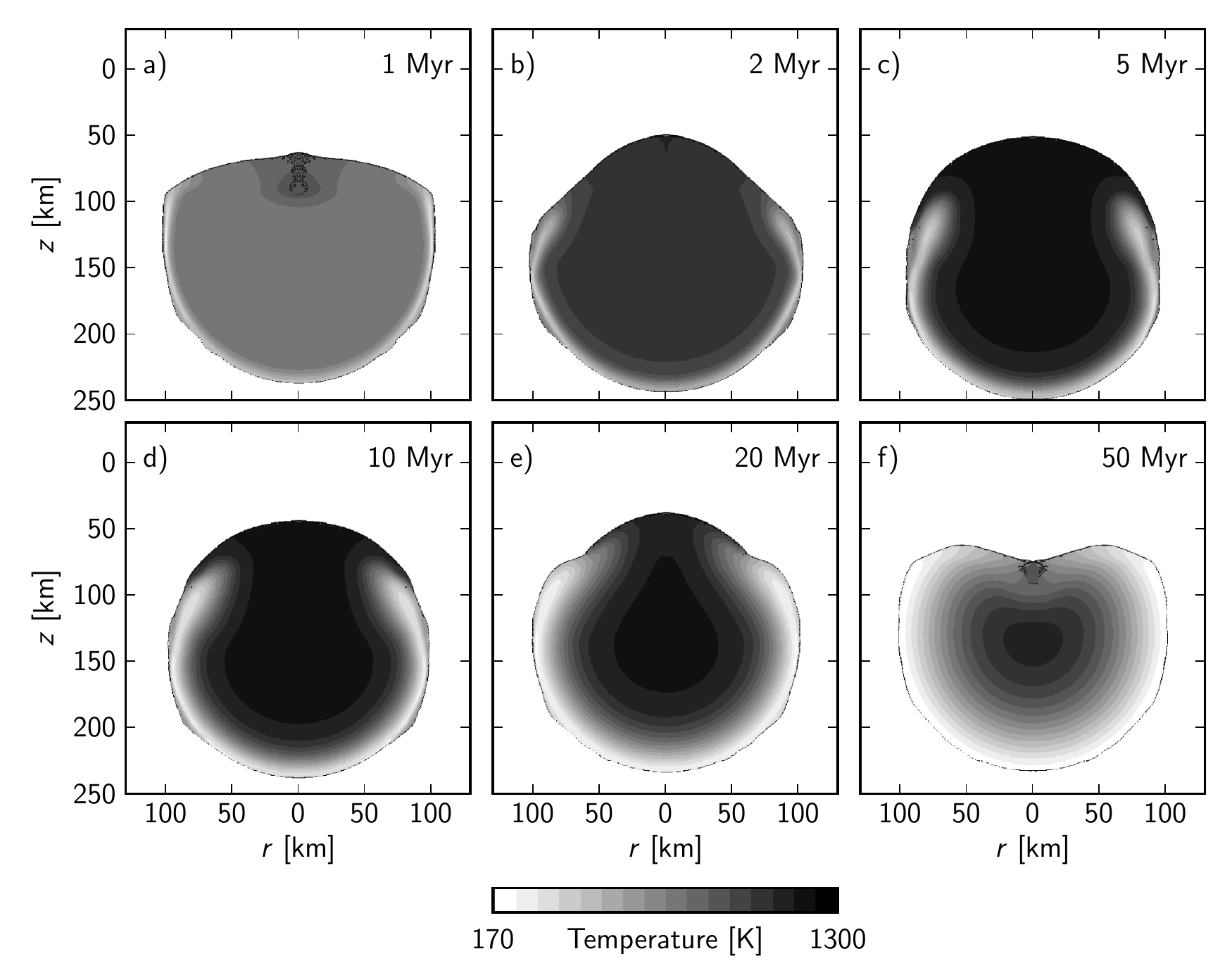}
\caption{\small Post-impact temperature distribution within planetesimals internally heated by radioactive decay of $^{26}$Al for a single impact that occurs at different times during thermal evolution. At early and late times, the impact occurs on a relatively cool planetesimal, resulting in a recognizable crater with a localized, impact-heated region at its center. In contrast, impacts that occur during the period of high internal temperatures ($\sim$2-20 Ma for this size planetesimal) result in uplift and overturn of deeply-derived, hot and thermally-softened material. From Ciesla et al. (2013)}
\end{figure*}

The rapid cooling of iron meteorites is a longstanding problem (e.g. Wood 1964) that could perhaps be solved if melted core material buried deep inside a differentiated asteroid could  be exhumed by this mechanism, so that the iron might cool rapidly. But from many simulations (Love and Ahrens 1996, and thereafter) it appears that an iron core, once buried, is almost impossible to blast out by impact overturning. Furthermore it would sink back to the core, cooling there instead of at the surface. So the above process, of `breaking open the pie crust', would affect mostly the thermodynamics of meteorites from the upper to mid-mantle of a melted planetesimal, not the core. 

It is possible, on the other hand, to achieve greatly accelerated cooling of iron planetesimals if we consider the exhumation of the projectile rather than the target. As noted above, planetesimal projectiles (but not targets) during SSCs can be ripped apart into beads of iron-rich objects during hit and run collisions involving larger embryos (Asphaug et al. 2006; Asphaug 2010). The mechanism of mantle stripping of hit and run projectiles is discussed in more detail in the chapter by Scott et al. (2015), including its implications for the history and evolution of the asteroid parent body population. 

While mantle stripping by hit and run has been simulated dynamically in a number of studies, the thermal consequences of abruptly exposing interior materials to space awaits the kind of detailed study that Ciesla et al. (2013) have given to large scale cratering. Asphaug et al. (2011) consider exhumation in a more specific context to explain the cooling rates of chondrules, that are formed according to their model by inefficient accretion of incompletely differentiated melted planetesimals. For the general case, it is apparent that the accelerated-cooling scenarios defined above for mega-cratering would have comparable, more accentuated aspects if applied to the aftermath of hit and run collisions, that typically rip off half the mantle of a differentiated body without producing shocks (Asphaug 2010). Yang et al. (2007) invoke hit and run mantle stripping to explain the rapid cooling rates of the IVA iron meteorites, in a scenario that requires almost complete mantle stripping to leave behind only a thin rind of insulating silicates. This would require repeated hit and run collisions while the body was still molten -- something that Asphaug and Reufer (2014) argue would have been probable for at least a few of the unaccreted bodies that became the progenitors of asteroids.

\section{Rheology and disruption}

One goal of current numerical modeling is a unified simulation of asteroid collision including fracture, crushing, dilatatancy, friction, and debris flow. The physical modeling borrows equally from dynamics, seismology, geomorphology, and impact cratering. Only recenty has it been possible to look at the additional effects of friction, porosity and cohesion, at least on a global scale, so this represents a frontier of computational geophysics.   

\begin{figure*}
\epsscale{1.5}
\plotone{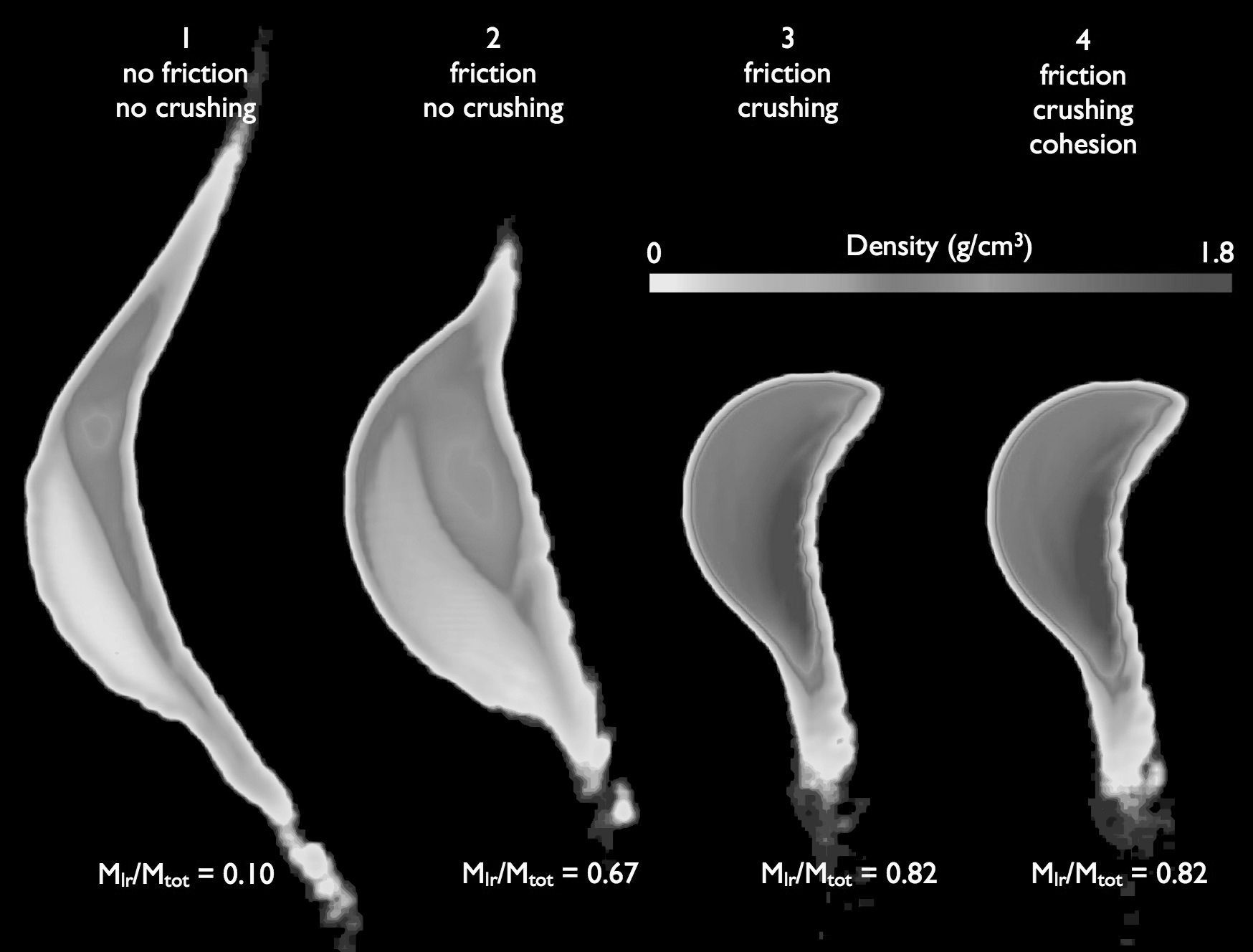}
\caption{\small Cross-sections of four SPH simulations of collisions between targets $R_t = 100$ km and projectiles $R_p = 27$ km, at relative velocity 3 km/s and 45$^\circ$ impact angle. Four different rheologies are investigated. The degree of disruption, the size of the largest remnant ($M_{lr}/M_{tot}$), and the increase in density strongly depend on the target properties. From Jutzi (2014).}
\end{figure*}

Jutzi (2014) used SPH to model colliding bodies in an exploration of the threshold for catastrophic disruption. Below we show a subset of those calculations: colliding asteroids that are the same size in each simulation ($R_t = 100$ km, $R_p = 27$ km), and the same mass (initial bulk density $\rho$ = 1.3 g/cm$^3$), and the same impact parameter ($v_{imp}=3$ km/s, $\theta=45^\circ$). What is different between the four simulations in Figure 6 are the material properties: 
\begin{enumerate}
\item no friction, no crushing, no cohesion
\item friction, no crushing, no cohesion 
\item friction and crushing, no cohesion 
\item friction, crushing and cohesion
\end{enumerate}

For the two cases with crushing, a crush curve with parameters for pumice is applied (Jutzi et al. 2009). For the case with cohesion, an internal friction $\mu_i$ = 1.5 and cohesion $Y_0$ = 100 MPa are applied, which is reduced to $\mu_d$ = 0.8 and $Y_m$ = 3.5 GPa following damage. The first case in Figure 6 is a fluid, the second is a fluid resisted by friction, the third is a crushable fluid resisted by friction, and the last is a cohesive porous body like pumice. 

Based on a suite of such models, Jutzi (2014) found the catastrophic disruption energy $Q^*_D$ to be $\sim5-10$ times higher when friction is included. The disruption threshold further increases by a factor $\sim 2-3$ when the energy dissipation by compaction (pore crushing) is taken into account. (Surprisingly, there is not much difference between the case with or without cohesion.) This implies that many events previously thought of as catastrophically disruptive, are more likely global scale impacts that leave behind a completely transformed asteroid, but do not disperse it. 

These simulations are consistent with previous models, that cohesion and tensile strength are negligible in potentially disruptive impacts involving bodies larger than a few 100 m diameter. But it is a misnomer to speak of this as the `strength-gravity transition', because gravity is not the controlling aspect for the response, either. For example, collisions involving partially melted silicate targets up to $\sim 1000$ km diameter may be limited by viscosity (Asphaug et al. 2006). For solid asteroids and planetesimals up to 100s of km diameter that have been  beaten into rubble, friction and compaction are expected to dominate, according to these calculations. Although gravity cannot be ignored, there may be no simple `gravity regime' for asteroid impacts. 

\section{Conclusions}

Global scale impacts on asteroids and their parent bodies are in many respects similar to the giant impacts that created the terrestrial planets, and there is great scientific value in making the connection. But at asteroid scales, we cannot ignore the internal stresses and their associated rheological responses -- the breaking of rock, the flow of rubble, the opening of faults, the seismological vibration -- in addition to the self-gravitational complexities. 

Gravity is subtle, and so are the other forces, and approximations or even small numerical errors can overwhelm the accurate calculation of an asteroid impact. In recent years computer modeling has begun to catch up with the physics, and progress on this front defines much of the modern science of global scale impacts. In the coming decade, we can look forward to ever more sophisticated simulations; for now it is too early to say whether the models are sufficient to be predictive.

Meanwhile, every new mission to an asteroid -- and in particular, every new mission to a different size or class of asteroid -- has revealed a solar system process not previously envisioned, and has turned our previous understanding of small bodies on its head. Flyby missions are of great value, as are the increasingly impressive radar shape models (Ostro et al. 1990) and adaptive optics models (see chapter by Marchis et al. 2015) obtained so far for dozens of NEAs (see chapter by Benner et al. 2015). While lacking the fidelity of a spacecraft flyby, radar images are sufficiently detailed to play a fundamental role in resolving the geophysical debates that have been summarized in this chapter.

Rendezvous missions provide a much more thorough examination of a smaller subset -- three asteroids so far, with three more on the near horizon: the sample return missions Hayabusa 2 and OSIRIS-REx to small NEOs, and the Dawn mission continuing to Ceres. Hayabusa 2 will conduct the first impact experiment on an asteroid, and although it is almost certainly too small of a scale to result in any of the global scale phenomena described above, it will no doubt shed light on some of the wide-open questions concerning microgravity impact physics at several meters scale. 

Efforts are ongoing to study the effect of a small spacecraft colliding into an asteroid using an LCROSS-like mission approach (Colaprete et al. 2010), which depending on the asteroid could cause global accelerations greater than local gravity, rearranging the global topography. Such a mission would elucidate the many factors that dictate whether a cratering event has global consequences or is local to one hemisphere -- an understanding which in turn will allow us to put the more readily attainable cratering observations from radar and flyby missions into context.

And the last front is meteorites, including the samples that are collected by missions. Meteorites have a way of presenting us with seemingly contradictory information, leading to a spectrum of contested hypotheses for their origin. But there is a chance, through collisional and thermal modeling and geochemistry, to understand their most basic physical properties and the processes they have experienced, from the formative collisions to the catastrophic disruptions that have populated the Main Belt and near-Earth space.

\bigskip

\centerline\textbf{ ACKNOWLEDGEMENTS:} EA acknowledges NASA PGG grant NNX13AR66G. GSC acknowledges STFC grant ST/J001260/1. MJ acknowledges the Ambizione program of the Swiss National Science Foundation. We are grateful to Tom Davison for helpful discussion and access to unpublished data, and acknowledge the careful work of two referees to improve the clarity and balance of the paper. 
\bigskip

\centerline\textbf{ REFERENCES}
\bigskip
\parskip=0pt
{\small
\baselineskip=11pt

\refs Agnor C. B., Canup, R. M. \& Levison, H. F. (1999) On the character and consequences of large impacts in the late stage of terrestrial planet formation. Icarus 142, 219-237
\refs Ahrens, T. J. \& O'Keefe, J. D. (1994) Impact-induced melting of planetary surfaces. In Dressler, B. O., Grieve, R. A. F., \& Sharpton, V. L., eds., Large Meteorite Impacts and Planetary Evolution, Geological Society of America Special Paper 283, 103-109
\refs Amsden, A., Ruppel, H., \& Hirt, C. (1980) SALE: A simplified ALE computer program for fluid flow at all speeds. Los Alamos National Laboratories Report, LA-8095, 101pp. 
\refs Anderson, C. E. (1987) An overview of the theory of hydrocodes. Int. J. Impact Engineering 5, 33-59
\refs Asphaug, E. \& Melosh, H. J. (1993) The Stickney impact of Phobos - A dynamical model. Icarus 101, 144-164
\refs Asphaug, E. (1997) Impact origin of the Vesta family. Meteoritics and Planetary Science 32, 965-980
\refs Asphaug, E., Ryan, E.V., \& Zuber, M.T. (2002) Asteroid interiors. In: Bottke, W.F., Cellino, A., Paolicchi, P. \& Binzel, R.P. (Eds.), Asteroids III. Univ. of Arizona Press, Tucson, pp. 463-484.
\refs Asphaug, E., Agnor, C. B. \& Williams, Q. (2006) Hit-and-run planetary collisions. Nature 439, 155-160
\refs Asphaug, E. (2008). Critical crater diameter and asteroid impact seismology. Meteoritics \& Planetary Science 43, 1075-1084
\refs Asphaug, E. (2009) Growth and Evolution of Asteroids. Annual Review of Earth and Planetary Sciences 37, 413-448
\refs Asphaug, E. (2010) Similar-sized collisions and the diversity of planets. Chemie der Erde / Geochemistry 70, 199-219
\refs Asphaug, E., Jutzi, M. \& Movshovitz, N. (2011) Chondrule formation during planetesimal accretion. Earth \& Planetary Science Letters 308, 369-379
\refs Asphaug, E., Ostro, S. J., Hudson, R. S., Scheeres, D. J. \& Benz, W. (1998) Disruption of kilometre-sized asteroids by energetic collisions. Nature 393, 437-441 
\refs Asphaug, E. \& Reufer, A. (2013) Late origin of the Saturn system. Icarus 223, 544-565
\refs Asphaug, E. and Reufer, A. (2014) Mercury and other iron-rich planetary bodies as relics of inefficient accretion. Nature Geoscience, published online July 6, 2014, doi: 10.1038/ngeo2189
\refs Belton, M. J. S. et al. (2007) The internal structure of Jupiter family cometary nuclei from Deep Impact observations: The ÔtalpsÕ or Ôlayered pileÕ model. Icarus 187, 332-344
\refs Benner et al. (2015) Radar observations of near-Earth and main belt asteroids. This volume. 
\refs Benz, W., \& Asphaug, E. (1994) Impact simulations with fracture. I - Method and tests. Icarus 107, 98-116
\refs Benz, W. \& Asphaug, E. (1995) Simulations of brittle solids using smooth particle hydrodynamics. Computer Physics Communications 87, 253-265
\refs Benz, W. \& Asphaug, E. (1999) Catastrophic disruptions revisited. Icarus 142, 5-20
\refs Benz, W. (2000) Low velocity collisions and the growth of planetesimals. Space Science Reviews 92, 279-294
\refs Blitz, C.,  Lognonn\`e, P., Komatitsch, D. \&  Baratoux, D. (2009) Effects of ejecta accumulation on the crater population of asteroid 433 Eros, Journal of Geophysical Research - Planets 114, E06006, 1-10
\refs Binzel, R. P. \& Xu, S. (1993) Chips off of asteroid 4 Vesta - Evidence for the parent body of basaltic achondrite meteorites. Science 260, 186-191
\refs Bizzarro, M., Baker, J. A., Haack, H. \& Lundgaard, K. L. (2005) Rapid Timescales for Accretion and Melting of Differentiated Planetesimals In- ferred from 26Al-26Mg Chronometry. The Astrophysical Journal 632, L41-L44
\refs Bowling, T. J., Johnson, B. C., Melosh, H. J., Ivanov, B. A., O'Brien, D. P., Gaskell, R., \& Marchi, S. (2013) Antipodal terrains created by the Rheasilvia basin forming impact on asteroid 4 Vesta. Journal of Geophysical Research: Planets 118 (9), 1821-1834
\refs Canup, R. M. (2004) Dynamics of Lunar Formation. Annual Review of Astronomy and Astrophysics 42, 441-475
\refs Carroll, M.M., \& Holt, A.C. (1972) Suggested modification of the $P$ - $\alpha$ model for porous materials. J. Appl. Phys. 43, 759-761.
\refs Ciesla F.J., Davison T. M., Collins G.S. \& O'Brien D.P.  (2013) Thermal consequences of impacts in the early solar system. Meteoritics \& Planetary Science 48, 2559-2576
\refs Clenet, H., Jutzi, M., Barrat, J.-A., Asphaug, E., Benz, W. \& Gillet, P. (2014) A deep crust-mantle boundary in the asteroid 4 Vesta. Nature 511, 303-306
\refs Colaprete, A. et al. (2010) Detection of Water in the LCROSS Ejecta Plume. Science 330, 463-467
\refs Collins, G. S., Melosh, H. J., \& Ivanov, B. A. (2004) Modeling damage and deformation in impact simulations. Meteorit. Planet. Sci. 39, 217--231
\refs Collins, G. S., Melosh, H. J. \& W\"unnemann, K. (2011) Improvements to the $\epsilon$-$\alpha$ porous compaction model for simulating impacts into high-porosity solar system objects. Int. J. Imp. Eng. 38, 434–-439.
\refs Collins, G. S., W\"unnemann, K., Artemieva, N. \& Pierazzo, E. (2013) Numerical modelling of impact processes, in Impact Cratering: Processes and Products 254-270 (Wiley-Blackwell).
\refs Cooper, H.F., Jr. \& Sauer, F.M. (1977) Crater-related ground motions and implications for crater scaling.  In Impact and Explosion Cratering (Roddy, D.J., Pepin, R.O. and Merrill, R.B., eds.), Pergamon Press, 1133-1163.
\refs Crawford, D.A., Taylor, P.A. Bell, R.L., \& Hertel, E.S. (2006) Adaptive mesh refinement in the CTH shock physics hydrocode. Russ. J. Phys. Chem. B, 25 (9) 85-90
\refs Davis, D. R., Durda, D. D., Marzari, F., Campo Bagatin, A. \& Gil-Hutton, R. (2002) Collisional Evolution of Small-Body Populations. In: Bottke, W.F., Cellino, A., Paolicchi, P. \& Binzel, R.P. (Eds.), Asteroids III. Univ. of Arizona Press, Tucson, pp. 545-558
\refs Davison T. M., Collins G. S., \& Ciesla F. J. (2010) Numerical modelling of heating in porous planetesimal collisions. Icarus 208:468-481.
\refs Davison, T. M., Ciesla, F. J. \& Collins, G. S. (2012) Post-Impact Thermal Evolution of Porous Planetesimals. Geochimica et Cosmochimica Acta 95, 252-269.
\refs Davison T. M., O'Brien D. P., Ciesla F. J., and Collins G. S. (2013) The early impact histories of meteorite parent bodies. Meteoritics \& Planetary Science 48:1894-1918
\refs Durda, D. D. et al. (2004). The formation of asteroid satellites in large impacts: results from numerical simulations. Icarus 170, 243-257
\refs Farinella, P., Paolicchi, P. \& Zappal\'a, V. (1982) The asteroids as outcomes of catastrophic collisions. Icarus 52, 409-433
\refs Flynn, G.J., (2005) Physical properties of meteorites and interplanetary dust particles: Clues of the properties of meteors and their parent bodies. Earth, Moon, and Planets, 95, 361-374. 
\refs Grady, D.E. \& Kipp, M.E. (1980) Continuum modelling of explosive fracture in oil shale. Int. J. Rock Mech. Min. Sci. \& Geomech. Abstr. 17, 147-157.
\refs Herrmann, W. (1969) Constitutive equation for the dynamic compaction of ductile porous materials. J. Appl. Phys. 40, 2490-2499.
\refs Holsapple, K.A. (2009) On the ``strength" of the small bodies of the solar system: A review of strength theories and their implementation for analyses of impact disruptions. Planetary and Space Science 57, 127-141
\refs Housen, K. R., Schmidt, R. M. \& Holsapple, K. A. (1983) Crater ejecta scaling laws - Fundamental forms based on dimensional analysis. Journal of Geophysical Research 88, 2485-2499
\refs Housen, K. R. \& Holsapple, K. A. (1990) On the fragmentation of asteroids and planetary satellites. Icarus 84, 226-253.
\refs Housen, K. R. \& Holsapple, K. A. (1999) Scale Effects in Strength-Dominated Collisions of Rocky Asteroids. Icarus 142, 21-33
\refs Housen, K. R. \& Holsapple, K. A. (2011) Ejecta from impact craters. Icarus 211, 856-875. 
\refs Ivanov B. A., Deniem D. \& Neukum G. (1997) Implementation of dynamic strength models into 2D hydrocodes: Applications for atmospheric breakup and impact cratering. Int. J. Impact Eng. 20, 411-430.
\refs Jackson, A. P. \& Wyatt, M. C. (2012) Debris from terrestrial planet formation: the Moon-forming collision. Monthly Notices of the Royal Astronomical Society 425, 657-679 
\refs Johansson, A. et al. (2015). New paradigms for asteroid formation. This volume
\refs Jutzi, M., Benz, W., Michel, P. (2008) Numerical simulations of impacts involving porous bodies. I. Implementing sub-resolution porosity in a 3D SPH hydrocode. Icarus 198, 242-255
\refs  Jutzi, M., Michel, P., Hiraoka, K, Nakamura, A.M. \& Benz, W. (2009) Numerical simulations of impacts involving porous bodies. II. Comparison with laboratory experiments. Icarus 201, 802=-813
\refs Jutzi, M. \& Asphaug, E. (2011) Mega-ejecta on asteroid Vesta. Geophysical Research Letters 38, 1102
\refs Jutzi, M., Michel, P., Benz, W. \& Richardson D. C. 2010. Fragment properties at the catastrophic disruption threshold: The effect of the parent body's internal structure. Icarus 207:54-65.
\refs Jutzi, M., Asphaug, E., Gillet, P., Barrat, J-A. \& Benz, W. (2013) The structure of the asteroid 4 Vesta as revealed by models of planet-scale collisions. Nature 494, 207-210
\refs Jutzi, M. (2014) SPH calculations of asteroid disruptions: The role of pressure dependent failure models. Planetary and Space Science, http://dx.doi.org/10.1016/j.pss.2014.09.012
\refs Jutzi, M. et al. (2015) Modeling collisions and impact processes. This volume
\refs Keil K., Stoeffler D., Love S. G. \& Scott E. R. D. (1997) Constraints on the role of impact heating and melting in asteroids. Meteoritics \& Planetary Science 32, 349-363
\refs Korycansky, D. G. \& Asphaug, E. (2006) Low-speed impacts between rubble piles modeled as collections of polyhedra. Icarus 181, 605-617
\refs Korycansky, D. G. \& Asphaug, E. (2009) Low-speed impacts between rubble piles modeled as collections of polyhedra, 2. Icarus 204, 316-329
\refs Kraus, R.G., Stewart, S.T., Swift, D.C., Bolme, C.A., Smith, R.F., Hamel, S., Hammel, B.D., Spaulding, D.K., Hicks, D.G., Eggert, J.H. \& Collins, G. W. (2012) Shock vaporization of silica and the thermodynamics of planetary impact events. Journal of Geophysical Research: Planets, Volume 117, Issue E9, DOI: 10.1029/2012JE004082
\refs Leinhardt, Z. M. \& Stewart, S. T. (2012) Collisions between gravity-dominated bodies. I. Outcome regimes and scaling laws. Astrophysical Journal 745, 79-106
\refs Leinhardt, Z. M. \& Stewart, S. T. (2009) Full numerical simulations of catastrophic small body collisions. Icarus 199 (2009) 542-559
\refs Love S. G. \& Ahrens T. J. (1996) Catastrophic impacts on gravity dominated asteroids. Icarus 124, 141-155
\refs Marchis, S. et al. (2015), this volume
\refs McGlaun, J.M., Thompson, S.L., \& Elrick, M.G. (1990) CTH: A 3-dimensional shock-wave physics code. Int. J. Imp. Eng., 10, 351-360
\refs McSween, H. J. et al. (2013) Composition of the Rheasilvia basin, a window into VestaÕs interior. J. Geophys. Res. 118, 335-346
\refs Melosh, H. J. (1989) Impact cratering: A geologic process. Oxford Monographs on Geology and Geophysics, No. 11, 253 pp.
\refs Melosh H. J., Ryan E. V., \& Asphaug E. (1992) Dynamic Fragmentation in Impacts: Hydrocode Simulation of Laboratory Impacts. J. Geophys. Res. 97(E9), 14735-14759.
\refs Melosh, H. J. \& Ivanov, B. A. (1999) Impact crater collapse. Annu. Rev. Earth Planet. Sci. 27, 385-415.
\refs Melosh, H.J. (2007) A hydrocode equation of state for SiO2. Meteoritics and Planetary Science, 42, 2079-2098.
\refs Michel, P., Benz, W. \& Richardson, D. C. (2003) Disruption of fragmented parent bodies as the origin of asteroid families. Nature 421, 608-611
\refs Michel, P. et al. (2015) Collisional formation and modeling of families. This volume
\refs Monaghan, J.J. (2012). Smoothed Particle Hydrodynamics and Its Diverse Applications. Ann. Rev. Fluid Mech. 44, 323-346
\refs Movshovitz, N., Asphaug, E. \& Korycansky, D. (2012) Numerical modeling of the disruption of comet D/1993 F2 Shoemaker-Levy 9 representing the progenitor by a gravitationally bound assemblage of randomly shaped polyhedra. Astrophys. J. 759, 93
\refs Murdoch, N. et al. (2015), this volume
\refs O'Brien, D.P. \& Greenberg, R. (2005) The collisional and dynamical evolution of the main-belt and NEA size distributions. Icarus 178, 179-212.
\refs Ostro, S. J. et al. (2002). Asteroid Radar Astronomy. In: Bottke, W.F., Cellino, A., Paolicchi, P. \& Binzel, R.P. (Eds.), Asteroids III. Univ. of Arizona Press, Tucson, pp.151-168
\refs Pierazzo E., Artemieva N., Asphaug E., Baldwin E. C., Cazamias J., Coker R., Collins G. S., Crawford D., Elbeshausen D., Holsapple K. A., Housen K. R., Korycansky D. G., \& W\"unnemann K. (2008) Validation of numerical codes for impact and explosion cratering. Meteoritics and Planetary Science 43, 1917-1938.
\refs Richardson, D. C., Leinhardt, Z., Melosh, H. J., Bottke, W. F., \& Asphaug, E. (2002). Gravitational aggregates: evidence and evolution. In Asteroids III (W.F. Bottke, Jr., A. Cellino, P. Paolicchi, and R.P. Binzel, eds.), University of Arizona Press, pp. 501-516
\refs Richardson, J. E., Melosh, H. J. \& Greenberg, R. (2004) Impact-induced seismic activity on asteroid 433 Eros: a surface modification process. Science 306, 1526-1529
\refs Russell et al. (2015). Vesta in the light of Dawn. This volume
\refs Ryan, E. V. \& Melosh, H. J. (1998) Impact Fragmentation: From the Laboratory to Asteroids. Icarus 133, 1-24 
\refs Safronov, V. S. (1972) Evolution of the protoplanetary cloud and formation of the Earth and planets. Jerusalem: Israel Prog. Sci. Translations, NASA TT F-677
\refs S\'anchez, P. \& Scheeres, D. J. (2011) Simulating asteroid rubble piles with a self-gravitating soft-sphere distinct element method model. Astrophysical Journal 727 (2), 120
\refs Scheeres, D. J. et al. (2015) Asteroid interiors and morphology. This volume.
\refs Schenk, P. et al. (2012) The geologically recent giant impact basins at VestaÕs south pole. Science 336, 694-697
\refs Scott, E. R. D. et al. (2015) Dynamical origin and impact history of differentiated meteorites and asteroids. This volume
\refs Sharp T. G. \& de Carli P. S. (2006) Shock effects in meteorites. In Meteorites and the Early Solar System II (eds. D. S. Lauretta and H. Y. McSween). University of Arizona Press, Tucson, pp. 653-677
\refs Stevenson, D. J. (1987) Origin of the moon - The collision hypothesis. Annual Review of Earth and Planetary Sciences 15, 271-315
\refs Stewart, S.T. \& Leinhardt, Z. M. (2012) Collisions between gravity-dominated bodies. II. The diversity of impact outcomes during the end state of planet formation.  Astrophysical Journal 751, 32-49
\refs Stickle, A. M., Schultz, P. H. \& Crawford, D. A. (2015) Subsurface failure in spherical bodies: A formation scenario for linear troughs on VestaÕs surface. Icarus 247, 18-34
\refs Thomas, P. C. \& Robinson, M. S. (2005) Seismic resurfacing by a single impact on the asteroid 433 Eros. Nature 436, 366-369
\refs Thompson, S.L. and Lauson, H.S. (1972) Improvements in the chart- D radiation hydrodynamic code III: revised analytical equation of state. Report SC-RR-710714, Sandia National Laboratories, Albuquerque, NM
\refs Tillotson, J. H. (1962) Metallic equations of state for hypervelocity impact. General Atomic Report GA-3216
\refs Weidenschilling, S. J. \& Cuzzi, J. N. (2006) Accretion dynamics and timescales: relation to chondrites. In Meteorites and the Early Solar System II, 473-485
\refs Wood, J. A. (1964) The cooling rates and parent planets of several iron meteorites. Icarus 3, 429-459
\refs Veverka, J. et al. (2000) NEAR at Eros: imaging and spectral results. Science 289, 2088-2097
\refs W\"unnemann, K., Collins, G.S. \& Melosh, H.J. (2006) A strain-based porosity model for use in hydrocode simulations of impacts and implications for transient crater growth in porous targets, Icarus, 180, 514-527
\refs Yang, J., Goldstein, J. I. \& Scott, E. R. D. (2007) Iron meteorite evidence for early formation and catastrophic disruption of protoplanets. Nature 446, 888-891

\end{document}